\documentclass[twocolumn]{aastex63}
\usepackage[utf8]{inputenc}
\usepackage{graphicx}
\usepackage{amssymb}
\usepackage{amsmath}
\usepackage{float}
\usepackage{multirow}
\usepackage{textcomp}
\usepackage{gensymb}
\usepackage{hyperref}
\usepackage{natbib}
\usepackage{comment}
\usepackage{systeme}
\usepackage[Symbol]{upgreek}
\usepackage{xcolor}
\definecolor{xlinkcolor}{cmyk}{1,1,0,0}

\shorttitle{X-ray and radio analysis of Abell 746}
\shortauthors{K. Rajpurohit et al.}

\begin{document}

\title{Abell 746: A highly disturbed cluster undergoing multiple mergers}

\correspondingauthor{Kamlesh Laxmi Rajpurohit}
\email{kamlesh.rajpurohit@cfa.harvard.edu}

\author[0000-0001-7509-2972]{K. Rajpurohit} 
\affil{Center for Astrophysics $|$ Harvard \& Smithsonian, 60 Garden Street, Cambridge, MA 02138, USA}
\affil{DIFA - Universit\'a di Bologna, via Gobetti 93/2, 40129 Bologna, Italy}
\affil{INAF-IRA, Via Gobetti 101, 40129 Bologna, Italy} 

\author[0000-0002-3754-2415]{L. Lovisari}
\affil{INAF-IASF Milano, Via A. Corti 12, 20133 Milano, Italy}
\affil{Center for Astrophysics $|$ Harvard \& Smithsonian, 60 Garden Street, Cambridge, MA 02138, USA}

\author[0000-0002-9325-1567]{A. Botteon}
\affil{INAF-IRA, Via Gobetti 101, 40129 Bologna, Italy} 

\author[0000-0003-2206-4243]{C. Jones}
\affil{Center for Astrophysics $|$ Harvard \& Smithsonian, 60 Garden Street, Cambridge, MA 02138, USA}

\author[0000-0002-9478-1682]{W. Forman}
\affil{Center for Astrophysics $|$ Harvard \& Smithsonian, 60 Garden Street, Cambridge, MA 02138, USA}

\author[0000-0002-5671-6900]{E. O'Sullivan}
\affil{Center for Astrophysics $|$ Harvard \& Smithsonian, 60 Garden Street, Cambridge, MA 02138, USA}

\author[0000-0002-0587-1660]{R. J. van Weeren}
\affil{Leiden Observatory, Leiden University, PO Box 9513, 2300 RA Leiden, The Netherlands}

\author{K. HyeongHan}
\affil{Department of Astronomy, Yonsei University, 50 Yonsei-ro, Seodaemun-gu, Seoul 03722, Republic of Korea}

\author[0000-0002-5068-4581]{A. Bonafede}
\affil{DIFA - Universit\'a di Bologna, via Gobetti 93/2, 40129 Bologna, Italy}
\affil{INAF-IRA, Via Gobetti 101, 40129 Bologna, Italy}

\author{M. J. Jee}
\affil{Department of Astronomy, Yonsei University, 50 Yonsei-ro, Seodaemun-gu, Seoul 03722, Republic of Korea}

\author[0000-0002-2821-7928]{F. Vazza}
\affil{DIFA - Universit\'a di Bologna, via Gobetti 93/2, 40129 Bologna, Italy}
\affil{INAF-IRA, Via Gobetti 101, 40129 Bologna, Italy}

\author{G. Brunetti}
\affil{INAF-IRA, Via Gobetti 101, 40129 Bologna, Italy}

\author[0000-0001-5966-5072]{H. Cho}
\affil{Department of Astronomy, Yonsei University, 50 Yonsei-ro, Seodaemun-gu, Seoul 03722, Republic of Korea}
\affil{ Center for Galaxy Evolution Research, Yonsei University, 50 Yonsei-ro, Seodaemun-gu, Seoul 03722, Republic of Korea}

\author[0000-0001-7058-8418]{P. Dom\'{i}nguez-Fern\'{a}ndez}
\affil{Center for Astrophysics $|$ Harvard \& Smithsonian, 60 Garden Street, Cambridge, MA 02138, USA}

\author{A. Stroe}
\altaffiliation{Clay Fellow}
\affil{Center for Astrophysics $|$ Harvard \& Smithsonian, 60 Garden Street, Cambridge, MA 02138, USA}

\author[0000-0002-4462-0709]{K. Finner}
\affil{IPAC, California Institute of Technology, 1200 E California Blvd., Pasadena, CA 91125, USA}

\author[0000-0002-3369-7735]{M. Br\"uggen}
\affil{Hamburger Sternwarte, Universit\"at Hamburg, Gojenbergsweg 112, 21029 Hamburg, Germany}

\author{J. M. Vrtilek}
\affil{Center for Astrophysics $|$ Harvard \& Smithsonian, 60 Garden Street, Cambridge, MA 02138, USA}

\author{L. P. David}
\affil{Center for Astrophysics $|$ Harvard \& Smithsonian, 60 Garden Street, Cambridge, MA 02138, USA}

\author{G. Schellenberger}
\affil{Center for Astrophysics $|$ Harvard \& Smithsonian, 60 Garden Street, Cambridge, MA 02138, USA}

\author{D. Wittman}
\affil{Department of Physics and Astronomy, University of California, Davis, CA 95616 USA}

\author{G. Lusetti}
\affil{Hamburger Sternwarte, Universit\"at Hamburg, Gojenbergsweg 112, 21029 Hamburg, Germany}

\author{R. Kraft}
\affil{Center for Astrophysics $|$ Harvard \& Smithsonian, 60 Garden Street, Cambridge, MA 02138, USA}

\author{F. de. Gasperin}
\affil{INAF-IRA, Via Gobetti 101, 40129 Bologna, Italy}
\affil{Hamburger Sternwarte, Universit\"at Hamburg, Gojenbergsweg 112, 21029 Hamburg, Germany}


\begin{abstract}
We present deep \textit{XMM-Newton}, Karl Jansky Very Large Array, and upgraded Giant Metrewave Radio Telescope observations of Abell 746, a cluster that hosts a plethora of diffuse emission sources that provide evidence for the acceleration of relativistic particles. Our new \textit{XMM-Newton} images reveal a complex morphology of the thermal gas with several substructures. We observe an asymmetric temperature distribution across the cluster: the southern regions exhibit higher temperatures, reaching $\sim$9\,keV, while the northern regions have lower temperatures ($\rm \leq4\,keV$), likely due to a complex merger. We find evidence of three surface brightness edges and one candidate edge, of which three are merger-driven shock fronts. Combining our new data with the published LOw-Frequency ARray observations has unveiled the nature of diffuse sources in this system. The bright northwest relic shows thin filaments and high degree of polarization with aligned magnetic field vectors. We detect a density jump, aligned with the fainter relic to the north. To the south, we detect high-temperature regions, consistent with shock-heated regions and density jump coincident with the northern tip of the southern radio structure. Its integrated spectrum shows a high-frequency steepening. Lastly, we find that the cluster hosts large-scale radio halo emission. The comparison of the thermal and nonthermal emission reveals an anticorrelation between the bright radio and X-ray features at the center. Our findings suggest that Abell 746 is a complex system that involves multiple mergers. 

\end{abstract}


\section{Introduction}
 \label{sec:intro}

In the standard hierarchical structure formation scenario, galaxy clusters are expected to form via a sequence of mergers with subclusters. In this process, the plasma contained within the subclusters collides, forming shocks and cold fronts. X-ray observations have revealed the evidence of these distinct X-ray surface brightness and temperature discontinuities \citep{Markevitch1997,Markevitch2005,Markevitch2001,Botteon2016b,George2017,vanWeeren2016a,Gennaro2019}. Merger shock fronts travel supersonically and are characterized by a temperature jump (also pressure) with higher temperatures downstream. Cold fronts are observed in both relaxed and merging clusters, characterized by a sudden drop in gas density while the gas temperature rises abruptly, maintaining pressure balance across the front \citep{Markevitch2001,Breuer2020}.

\setlength{\tabcolsep}{11pt}
\begin{table*}[!htbp]
\caption{Observational overview: VLA, uGMRT, and LOFAR observations.}
\begin{center}
\begin{tabular}{ l  c  c c c  c c}
  \hline  \hline  
\multirow{1}{*}{}& \multicolumn{2}{c}{VLA L-band} & \multirow{1}{*}{ uGMRT Band\,4}  &\multirow{1}{*}{LOFAR HBA$^{\ast}$}   \\  
 \cline{2-3} 
& B configuration & C configuration & & & \\
\hline
Frequency range&1-2\,GHz &1-2\,GHz&550-750\,MHz &120-169 MHz\\ 
Channel width &1\,MHz &1\,MHz & 97.7\,kHz &12.2\,kHz\\ 
Correlations &full Stokes&full Stokes&RR and LL&full Stokes\\
No of channels &1024&1024&2000 & 64 &\\ 
On source time &8\,hrs&6\,hrs &5.5\,hrs &8\,hrs \\
\hline 
\end{tabular}
\end{center}
{Notes. VLA observations were recorded in 16 spectral windows (64 channels in each spectral window); $^{\ast}$Observation presented in \cite{Botteon2022a}  }
\label{Tabel:obs}
\end{table*} 

Mergers between galaxy clusters can also be detected in the radio band through synchrotron emission originating from steep spectrum \footnote{We define the spectral index, $\alpha$, so that $S_{\nu}\propto\nu^{\alpha}$, where $S$ is the flux density at frequency $\nu$.} ($\alpha\leq-1$) diffuse sources that are directly associated with the intracluster medium (ICM). Recent rapid improvements in radio
telescope sensitivity and resolution are revealing unprecedented structures,  previously undetected, and complexity in the ICM \citep[e.g.,][]{Bonafede2021,Brienza2021,Botteon2022a,Knowles2022,Rajpurohit2022b,Rudnick2022,Ramatsoku2020,deGasperin2022}. Classically these structures include 1) radio relics, which are produced by merger-induced shock waves, 2) radio halos, produced by turbulence, and 3) radio emission from aged cosmic-ray electrons (CRe) which can be re-energized by several processes, likely related to the dynamics of the ICM  \citep[for theoretical and observational reviews, see ][]{Brunetti2014,vanWeeren2019}.

Radio relics are elongated radio sources, typically found in the outskirts of merging galaxy clusters, and they often show irregular surface brightness and filamentary morphology \citep{Owen2014,vanWeeren2017b,Rajpurohit2018,Gennaro2018,Rajpurohit2022b,deGasperin2022}. One notable characteristic of relics is strong polarization at GHz frequencies and magnetic field vectors along the merger axis of the relic \citep[e.g.,][] {Bonafede2012,vanWeeren2010,Kierdorf2016,Stuardi2019,Rajpurohit2020b,Loi2020,Rajpurohit2022a}. Relics are believed to be produced by the Diffusive Shock Acceleration (DSA) mechanism \citep{Blandford1987,Drury1983,Dolag2000,Hoeft2007}. However, this model suffers from the efficiency problem, that is, the energy dissipated by merger shocks is not enough to reproduce the luminosity of the relics via DSA if particles are accelerated directly from the thermal pool \citep{vanWeeren2016a,Botteon2020a}. 

As a remedy, shock re-acceleration or multiple-shock scenarios have been proposed \citep{Kang2016a,Kang2016b,Kang2021,Inchingolo2021}. In the first scenario,  a shock front re-accelerates electrons via DSA from an existing population of relativistic electrons \citep{Markevitch2005,Kang2016a}. There are a few examples that appear to show a connection between relics and lobes/tails of active galactic nuclei (AGN), as a possible source of fossil electrons \citep[e.g.,][]{Bonafede2014,vanWeeren2017a,Stuardi2019,Shimwell2015,HyeongHan2020}. In the second scenario, the relics are generated by multiple shocks induced in the turbulent ICM. Both shock re-acceleration and multiple-shock scenarios could alleviate the efficiency problem \citep[e.g.][]{2013MNRAS.435.1061P,Kang2021,Inchingolo2021}.

Unlike relics, radio halos are centrally located and extend throughout the cluster's volume. They are found in merging clusters and often show a correlation with the X-ray emission, suggesting a strong link between thermal and nonthermal plasma components \citep[e.g.,][]{Govoni2001a,vanWeeren2016a,deGasperin2020,Bonafede2022,Botteon2020a,Rajpurohit2021b,Rajpurohit2021c,Rajpurohit2023}. The turbulent re-acceleration models provide the most promising explanation for their origin, where cosmic-ray electrons (CRe) become radio-emitting after undergoing re-acceleration through turbulence injected into the ICM during cluster mergers \citep{Brunetti2001,Fujita2003,Cassano2005,Brunetti2011,Miniati2015, Pinzke2017}. 

\begin{figure*}[!thbp]
    \centering
    \includegraphics[width=0.9\textwidth]{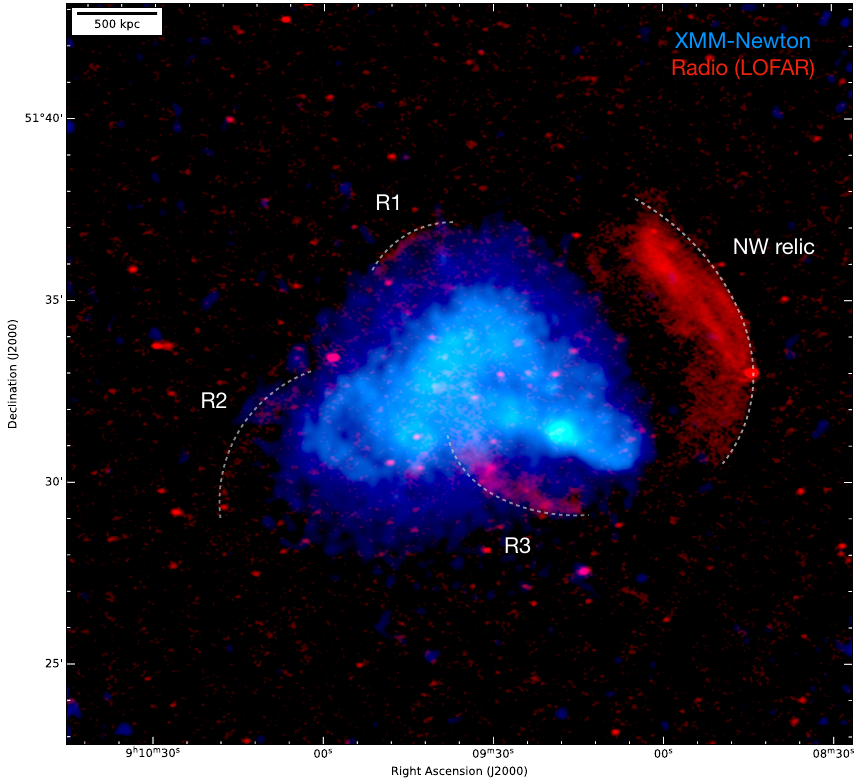}
    \vspace{-0.2cm}
 \caption{X-ray and radio overlay of Abell 746. The intensity in blue shows 
the {\it XMM-Newton} emission in the 0.7-2.0 keV band. The intensity in red shows the radio emission observed with LOFAR at a central frequency of 144\,MHz. The presence of four relics (R1, R2, R3, and NW relic) and the halo suggests a complex caused by multiple mergers. R2 and the halo are not clearly visible at this resolution, though, due to their low surface brightness, they were thus recovered better in the low-resolution images; see Figure\,\ref{high_res} right panels. The LOFAR image properties are given in Table\,\ref{imaging}, IM1. }
      \label{rgb}
\end{figure*}

Abell 746 (also known as PSZ2 G166.62+42.13; RXC J0909.3+5133), located at a redshift of $z = 0.23225$ \citep{Planck2016}, is a little-studied galaxy cluster. It has an X-ray luminosity of $\rm 3.2\times10^{44} erg\,s^{-1}$ (0.1-2.4 keV), a gas temperature of $4.6\pm0.3$ keV, and the mass of $M_{500}=5.34^{+0.39}_{-0.40}\times10^{14} M_{\odot}$\citep{Planck2016}. Despite the low X-ray luminosity and mass, the cluster contains multiple diffuse radio sources, which are tentatively classified as radio relics \citep[NW, R1, R2, and R3; see Figure\,\ref{rgb} for labeling;][]{vanWeeren2011a,Botteon2022a}. Recent low-frequency observations also hinted at the detection of a possible radio halo. \citep{Botteon2022a}. The presence of these features suggests that a major and/or multiple merger has occurred.

In this paper, we present the results of the combined \textit{XMM-Newton} and radio analysis of the galaxy cluster Abell 746. The new radio observations were performed with the upgraded Giant Metrewave Radio Telescope (uGMRT) at Band4 (550-750 MHz) and the Karl G. Jansky Very Large Array (VLA) at L-band (1-2\,GHz). We also used published LOw-Frequency ARray \citep[LOFAR;][]{Haarlem2013} High Band Antenna (HBA) observations \citep{Botteon2022a}.  

The layout of this paper is as follows. In Section\,\ref{observations}, we present an overview of the observations and the reduction of data. The new X-ray and radio images are presented in Section\,\ref{results}. The results obtained are described in Sections\,\ref{results} to \ref{origin}, followed by a summary in Section\,\ref{summary}.

Throughout this paper, we adopt a flat $\Lambda$CDM cosmology with $H_{\rm{ 0}}=69.6$ km s$^{-1}$\,Mpc$^{-1}$, $\Omega_{\rm{ m}}=0.286$, and $\Omega_{\Lambda}=0.714$ \citep{Wright2006}. At the cluster's redshift, $1\arcsec$ corresponds to a physical scale of 3.73\,kpc.


\section{Observations and data reduction}
\label{observations}

\subsection{XMM-Newton}
Abell 746 was recently observed four times (ObsIDs: 0902630101, 0902630201, 0902630301, 0902630601) by {\it XMM-Newton} (PI: C. Jones) with a total exposure time of $\sim$ 184 ks. Given that the last pointing is significantly shorter (i.e., $\sim$10 ks) than each of the other three pointings (i.e., $>$55 ks), we did not include it in the current analysis.

\setlength{\tabcolsep}{12.0pt}   
 \begin{table*}[!htbp]
\caption{Image properties}
\begin{center}
\begin{tabular}{c c c r c c c r}
\hline\hline
   & Name & Restoring Beam & Robust  & \textit{uv}-cut & \textit{uv}-taper & RMS noise\\ 
&&&parameter&&&$\upmu\,\rm Jy\,beam^{-1}$\\
\hline
  \hline 
&IM1&$8.0\arcsec \times 5.0\arcsec$&$-0.5$&$-$&$-$&70\\
LOFAR HBA&IM2&$7.0\arcsec \times 7.0\arcsec$&$-0.5$&$ \geq\rm0.1\,k\uplambda$&$-$&75\\
(120-168\,MHz)&IM3&$20\arcsec \times 20\arcsec$&$-0.5$&$-$&10\arcsec&130\\
&IM4&$20\arcsec \times 20\arcsec$&$-0.5$&$  \geq\rm0.1\,k\uplambda$&15\arcsec&140\\
\hline   
 &IM5&$5.9\arcsec \times 3.2\arcsec$&0.0&$-$&&7\\
 uGMRT Band\,4 &IM6&$7.0\arcsec \times 7.0\arcsec$&$-0.5$&$ \geq\rm0.1\,k\uplambda$&$2\arcsec$&11\\
 (550-750\,MHz)&IM7&$20\arcsec \times 20\arcsec$&$0.0$&$ -$&12\arcsec&14\\
 &IM8&$20\arcsec \times 20\arcsec$&$-0.5$&$ \geq\rm0.1\,k\uplambda$&15\arcsec&19\\
\hline   
&IM9&$9.5\arcsec \times 8.4\arcsec$&0.0&$-$&$-$&5\\
VLA L-band &IM10&$7.0\arcsec \times 7.0\arcsec$&$$-0.5$$&$\geq\rm0.1\,k\uplambda$&10\arcsec&7 \\
(1-2 GHz)&IM11&$20\arcsec \times 20\arcsec$&$0.0$&$-$&15\arcsec&10 \\
&IM12&$20\arcsec \times 20\arcsec$&$-0.5$&$ \geq\rm0.1\,k\uplambda$&15\arcsec&12 \\
 
 \hline   
 \end{tabular}
 \end{center}
{Notes. Imaging was always performed in {\tt WSCLEAN} using {\tt multiscale} and with {\tt Briggs} weighting scheme. Primary beam correction was performed in \textit{CASA}.  }
\label{imaging}
\end{table*}

Observation data files were processed with the {\it XMM-Newton} Science Analysis System (SAS) v20.0.0. We generated calibrated event files from raw data by running the tasks {\it emchain} and {\it epchain}. Then we excluded all events with PATTERN$>$12 for MOS data and with PATTERN$>$4 for pn data, and we followed the standard procedures for bright pixel and hot column removal (by applying the expression FLAG==0) and pn out-of-time correction. The data were cleaned for periods of high background induced by solar flares using the XMM-ESAS tools
{\it mos-filter} and {\it pn-filter}. The total exposure times after cleaning are 156.8  ks for MOS1, 158.1 ks for MOS2, and 104.2 ks for pn.  Additionally, we also excluded from the analysis all the CCDs in the so-called `anomalous state' (see \citealt{Kuntz2008} for more details). Point sources, identified by running the task {\it edetect-chain}, were excluded from the analysis. All the background event files were cleaned by applying the same PATTERN selection, flare rejection criteria, and point-source removal as used for the observation events.

All X-ray images presented in this work were obtained in the 0.7-2\,keV band with a binning of 40 physical pixels (corresponding to  2 arcsec) and by refilling the point-source regions using the CIAO task {\it dmfilth}.  The surface brightness profiles discussed in Section\,\ref{profiles} are extracted from a point source subtracted X-ray image with no refilling of point sources. 

\subsection{VLA}
We observed the cluster with the VLA in the L-band (1-2 GHz) in 2022 and 2023 in C and B configurations (PI: C. Jones), respectively. For observational details, see Table\,\ref{Tabel:obs}. All four polarization products (RR, RL, LR, and LL) were recorded. In the L-band, for both C and B configurations, 3C147 was included as the primary calibrator, observed for 5-10 minutes at the start of each observing run. The radio source J0834$+$5534 was included as a phase calibrator and 3C286 was a polarization calibrator.

The data were calibrated and imaged with the Common Astronomy Software Applications \citep[$\tt{CASA}$;][]{McMullin2007,casa2022} package. Data obtained from C and B configurations were calibrated separately, but in the same manner. The data reduction steps consisted of Hanning smoothing followed by RFI inspection. The bad data were flagged using ${\tt tfcrop}$ mode from the ${\tt flagdata}$ task. For target scans, we performed additional flagging using ${\tt AOFlagger}$ \citep{Offringa2010}. After this, we determined and applied elevation-dependent gain tables and antenna offset positions. We corrected the bandpass using 3C\,147. This prevents the flagging of good data due to the bandpass roll-off at the edges of the spectral windows. 

We used the L-band 3C147 and 3C286 models provided by the ${\tt CASA}$ software package and set the flux density scale according to \cite{Perley2013}. An initial phase calibration was performed using both calibrators over a few channels per spectral window. The antenna delays and bandpass response were determined using 3C147, followed by the gain calibration.

For polarization, the leakage response was determined using the unpolarized calibrator 3C147. Finally, the absolute position angle in the sky (the R-L phase difference) was corrected using the polarized calibrator 3C286. All calibration solutions were applied to the target field. The resulting calibrated data were averaged by a factor of 4 in frequency per spectral window to perform rotation measure synthesis (RM-synthesis).

\begin{figure*}[!thbp]
    \centering
    \includegraphics[width=1.0\textwidth]{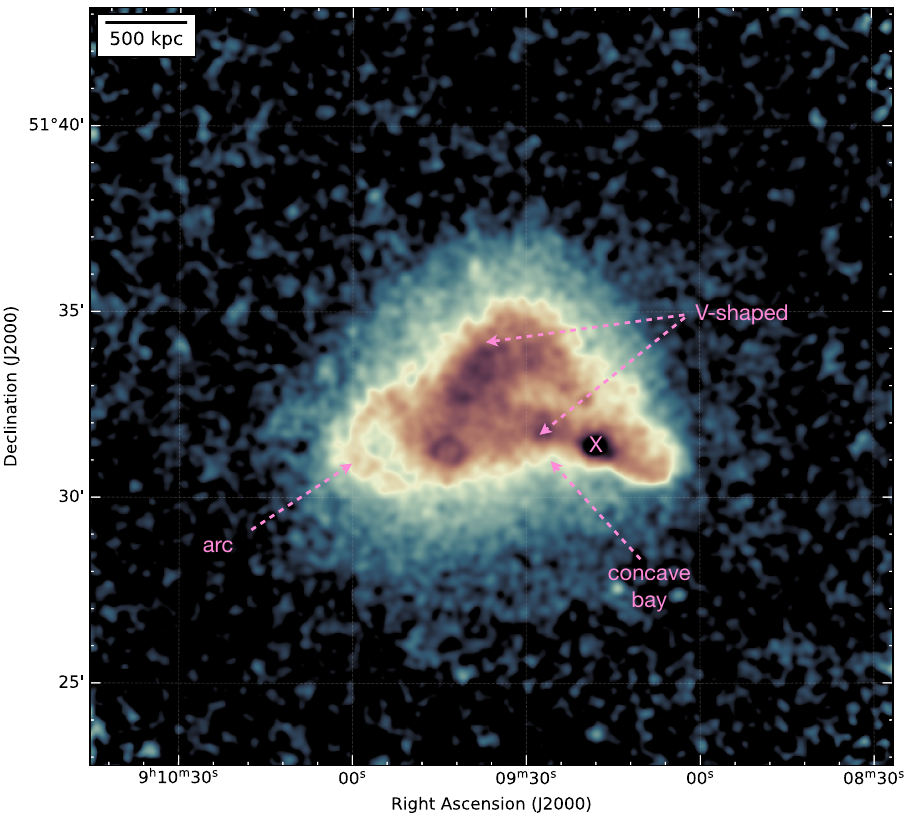}
        \vspace{-0.9cm}
 \caption{\textit{XMM-Newton} surface brightness map. The $0.7{-}2.0$\,keV band image was exposure-corrected and smoothed with a Gaussian kernel of 6\arcsec \,FWHM. All point sources are removed, with the exception of the one indicated by a cross mark. The image reveals multiple substructures, indicating the highly disturbed state of the cluster.}
      \label{XMM}
  \end{figure*}

After initial calibration and flagging, several rounds of self-calibration were performed further to refine the calibration for each individual data set. Imaging were done employing the W-projection algorithm in CASA \citep{Cornwell2008}. Clean masks were used for each imaging step. These masks were made using the PyBDSF source detection package \citep{Mohan2015}. The spectral index and curvature were taken into account during deconvolution using $\tt{nterms}=3$ \citep{Rau2011}.

After self-calibration, the C and B configurations data were combined. Final imaging of the combined data sets was done in {\tt WSClean} \citep{Offringa2014}, using Briggs weighting and employing the wideband and multiscale algorithms. The images were corrected for the primary beam attenuation in CASA.

\subsection{uGMRT}
The cluster was observed with the upgraded GMRT in Band\,4 (project code: 40\_025, PI: A. Botteon) using the GMRT Wideband Backend (GWB) covering a frequency range of 550-750\,MHz. The observations were carried out on May 6, 2021. In Table\,\ref{Tabel:obs}, we summarize the observations. Sources 3C48 and 3C147 were included as flux and phase calibrators.

The wideband GMRT data were processed using the Source Peeling and Atmospheric Modeling \citep[$\tt{SPAM}$;][]{Intema2009} pipeline{\footnote{\url{http://www.intema.nl/doku.php?id=huibintemaspampipeline}}}. For details about the main data reduction steps, we refer to \cite{Rajpurohit2021c}. In summary, the data were first split into six sub-bands. The flux densities of the primary calibrators were set according to \citet{Scaife2012}. Following flux density scale calibration, the data were averaged, flagged, and corrected for bandpass. We started self-calibration with a global sky model obtained with the GMRT narrow-band data. To produce deep full continuum images, the calibrated sub-bands were combined. The final deconvolution was performed in {\tt WSClean} using $\tt{multiscale}$ and $\tt{Briggs}$ weighting.

\subsection{LOFAR}
Abell 746 was observed with LOFAR HBA as part of the LOFAR Two-metre Sky Survey  \citep[LoTSS;][]{Shimwell2017,Shimwell2019,Shimwell2022}. The observations were conducted in HBA dual inner mode. To summarize, data reduction and calibration were performed with the LoTSS DR2 pipeline \citep{Tasse2020} followed by a final ``extraction+self-calibration" scheme \citep{vanWeeren2020}. Since Abell 746 is a PSZ2 cluster in the LOFAR DR2 data release \citep{Botteon2022a}, we used the same calibrated data. For a detailed description of the observation and data reduction, we refer to \cite{Botteon2022a}.

\subsection{Flux density scale}
The overall flux scale for all observations (LOFAR, uGMRT, and VLA) was verified by comparing the spectra of compact sources within the field of view between 144\,MHz and 2\,GHz. The uncertainty in the flux density measurements was estimated as
\begin{equation}
\Delta S =  \sqrt {(f \cdot S)^{2}+{N}_{{\rm{ beams}}}\ (\sigma_{{\rm{rms}}})^{2}},
\end{equation}
where $f$ is the absolute flux density calibration uncertainty, $S$ is the flux density, $\sigma_{{\rm{ rms}}}$ is the RMS noise, and $N_{{\rm{beams}}}$ is the number of beams. We assumed absolute flux density uncertainties of 10\% for LOFAR HBA data \citep{Shimwell2022}, 5\% for uGMRT Band\,4, and 2.5\% for the VLA data \citep{Perley2013}.  

\begin{figure*}[!thbp]
    \centering
    \includegraphics[width=0.90\textwidth]{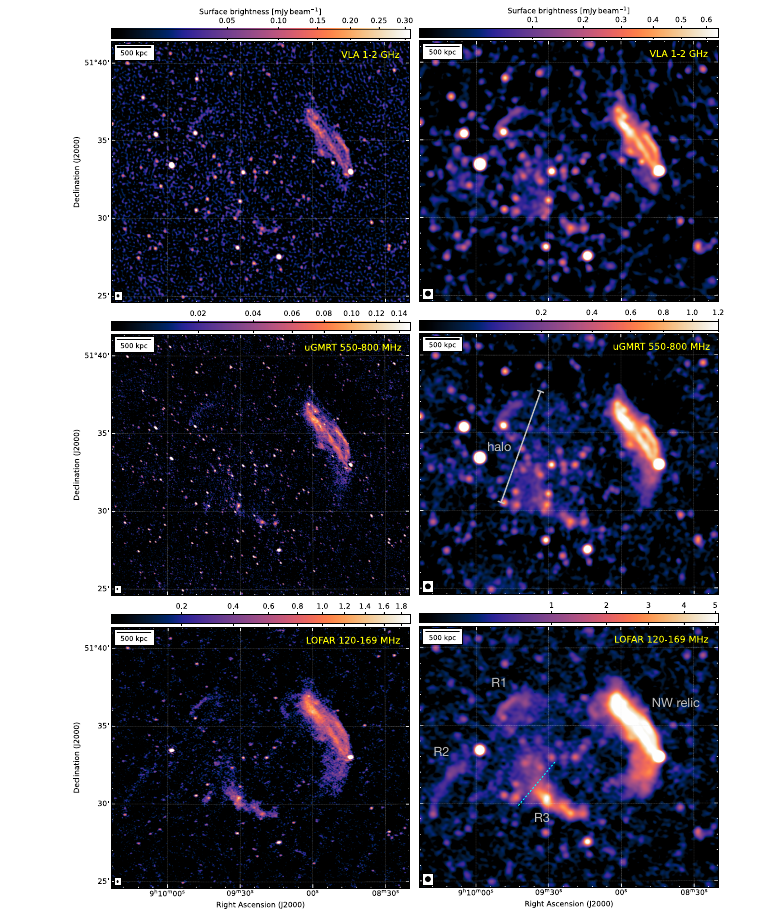}
    \vspace{-0.3cm}
 \caption{High (left) and low (right) resolution VLA  L-band (top), uGMRT Band4 (middle), and LOFAR  HBA (bottom) images of Abell 746 in square root scale. All low-resolution images are created at a common 20\arcsec\ resolution. The radio surface brightness unit is mJy\,$\rm beam^{-1}$. The beam size is indicated in the bottom left corner of each image. The images reveal a low surface halo emission and highlight the morphology of the diffuse emission sources as a function of observing frequency.  The dashed line shows the separation between R3 and the halo. The image properties are given in Table\,\ref{imaging}; IM1, IM5, IM9, IM3, IM7, and IM11.}
      \label{high_res}

  \end{figure*}  

The flux density values and the overall extent of the radio emission are extracted/measured, unless specified otherwise, in regions where the emission is $\geq3\sigma$. All output images are in the J2000 coordinate system and are corrected for primary beam attenuation.


\section{Results: X-ray and Radio Emission}
\label{results}

\subsection{X-ray Emission}

In Figure\,\ref{XMM}, we present the combined, background-subtracted, exposure-corrected, Gaussian-smoothed (6\arcsec) $0.7-2.0$ keV \textit{XMM-Newton} surface brightness map. All compact sources are subtracted except for a bright source, shown with cross marks. The disrupted X-ray morphology in the center, featuring several substructures, is clearly visible. 

The X-ray emission from the cluster is asymmetrical, with a roughly triangular morphology, see Figure\,\ref{XMM}. The central bright region of the cluster appears elongated along the southeast to northwest direction. The cluster does not exhibit a single surface brightness concentration instead, it reveals multiple distinct concentrations.  Similar to the distribution observed in other complex clusters such as MACS\,J0416.1-2403 \citep{Ogrean2015} and MACS\,J0717.5+3745 \citep{vanWeeren2017a}, it shows distinct features in the inner region. Moreover,  the cluster X-ray center cannot be identified by its morphology. The brightest part of the ICM consists of a ``V-shaped" structure which suggests some ongoing dynamical process in this cluster. A similar type of structure is seen in the multi-merger cluster MACS\,J0717.5+3745 \citep{vanWeeren2017a}.

\begin{figure*}[!thbp]
\centering
\includegraphics[width=0.49\textwidth]{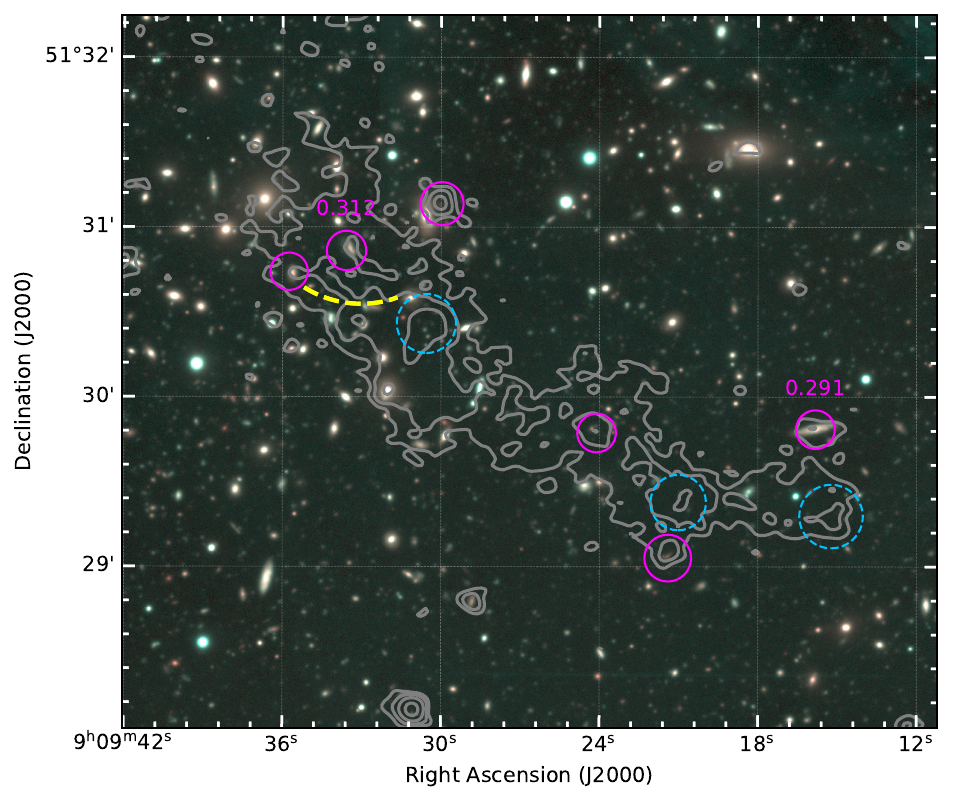}
\includegraphics[width=0.47\textwidth]{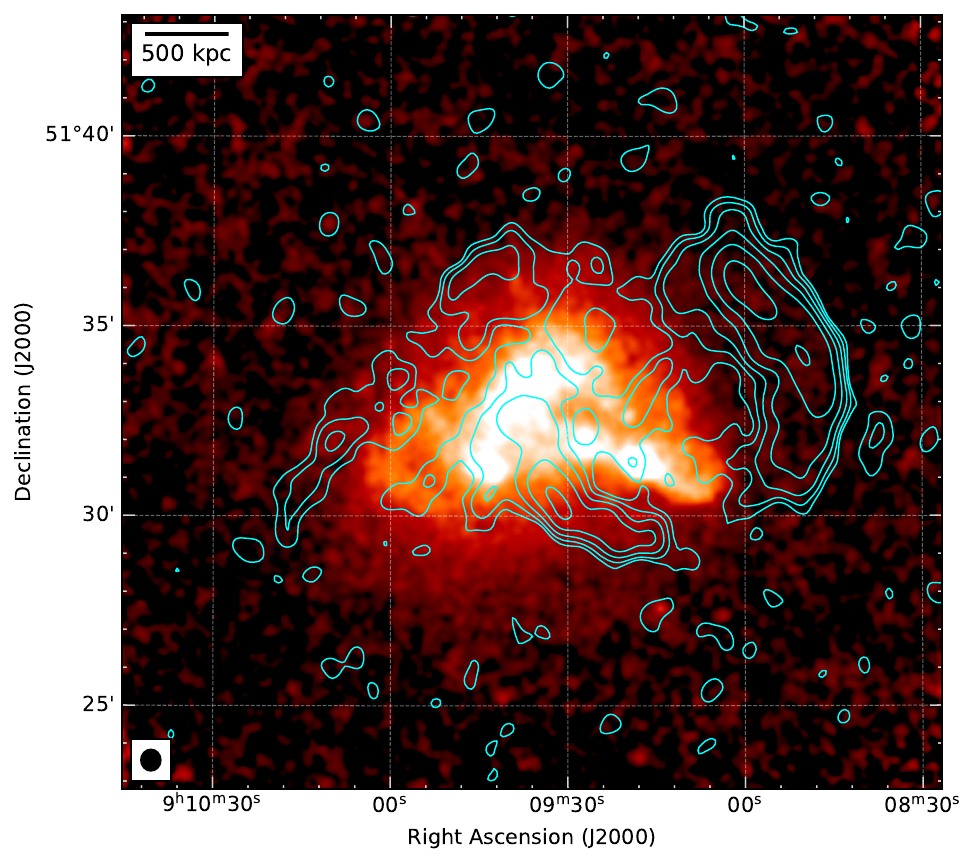}
    \vspace{-0.2cm}
 \caption{ \textit{Left}:  Subaru/Hyper-Suprime Cam zoom-in view of R3 \citep{Kim2023} overlaid with LOFAR 144\,MHz 7\arcsec\ resolution radio contours. For image properties, see Table\,\ref{imaging},\,IM2. The magenta circles mark the point sources embedded within R3 which have optical counterparts (with available redshift), while the cyan dashed circles represent compact regions without optical counterparts. The newly detected shock front position is indicated by the yellow dashed lines. \textit{Right}: \textit{XMM-Newton} image overlaid with 144\,MHz LOFAR contours.  The radio contours are from the 20\arcsec\ resolution image (point source subtracted) and drawn at  $[1, 2, 4, 8 ...]\times 3.0\sigma_{\rm rms}$ where ${\rm rms=130\,\mu Jy\,beam^{-1}}$.}
\label{radioxray}
\end{figure*}   

\setlength{\tabcolsep}{10pt}
\begin{table*}[htp]
\caption{Properties of the diffuse radio sources in the cluster Abell\,746.}
\begin{center} 
\begin{tabular}{*{8}{c}}
\hline \hline
\multirow{1}{*}{Source} &\multirow{1}{*}{LOFAR} &\multirow{1}{*}{uGMRT} &\multirow{1}{*}{VLA} &\multirow{1}{*}{$\rm LLS^{\dagger}$} & \multirow{1}{*}{$\alpha$}$^{\star}$& \multicolumn{2}{c}{Radio Power} \\
 \cline{7-8} 

& $S_{\rm144\,MHz}$ &${S_{\rm650\,MHz}}$&${S_{\rm1.5\,GHz}}$ & & & $P_{1.5\,\rm GHz}$& $P_{150\,\rm MHz}$\\
  & (mJy) & (mJy) & (mJy) & (Mpc) & &$(\rm 10^{24}\, W\,Hz^{-1}$)& $(\rm 10^{25}\,W\,Hz^{-1}$)\\
\cline{7-8}
  \hline 
NW & $447\pm60$&$70\pm10$ & $23\pm3$&$\sim1.8$ & $-1.26\pm0.04$& $3.9$&$7.7$\\
R1& $22\pm5$& $3.0\pm0.4$& $0.9\pm0.2$&$\sim0.38$& $-1.36\pm0.07$& $0.16$&$0.38$\\
R2  & $14\pm2$ &-&-&$\sim0.90$& -& -&-\\
R3  & $57\pm10$& $5.2\pm0.6$& $0.6\pm0.1$&$\sim0.95$ &-& $0.17$& $1.1$\\
Halo & $32\pm6$&$3.4\pm0.6$ & $1.2\pm0.3$  &$\sim1.0$ &  $-1.48\pm0.10$& &\\
\hline 
\end{tabular}
\end{center} 
{Notes. Flux densities were extracted from images created with ${\tt robust}=-0.5$ and a \textit{uv}-cut of $0.1k\lambda$. The image properties are given in Table\,\ref{imaging}, IM4, IM8, and IM14. The regions where the flux densities were extracted are indicated in the Figure\,\ref{regions}. Absolute flux density scale uncertainties are assumed to be 10\% for LOFAR, 5\% for the uGMRT Band4, and 2.5\% for the VLA L-band data. $^{\dagger}$The LLS measured at 144\,MHz; $^{\star}$The integrated spectral index obtained by fitting a single power-law fit. }
\label{Tabel:Tabel2}   
\end{table*}    

The X-ray emission to the southwest shows a clear ``concave-bay" morphology which extends from southwest to southeast, see Figure\,\ref{XMM}. This structure is apparently similar to the one reported in Abell 2256, known as `shoulder' cold front \citep{Sun2002,Breuer2020,Ge2020}. To the southeast, there is a surface brightness decrement, followed by a curved X-ray substructure, labeled as the arc.

\subsection{Radio emission}
In Figure\,\ref{high_res}, we show the high (left panel) and low-resolution (right panel) radio images of the cluster using VLA, uGMRT, and LOFAR observations.  The high radio surface brightness of the northwest relic allowed its detection at all observed frequencies in the high-resolution images, while other features are not prominently visible due to their lower surface brightness. The northwest relic has the largest linear size (LLS) of 1.1 Mpc, 1.4 Mpc, and 1.8 Mpc at 1.5 GHz, 650 MHz, and 144 MHz, respectively. The NW relic's southwest tip displays diffuse emission that extends progressively towards lower frequencies, see Figure\,\ref{high_res} right panel. Additionally, the northwest relic appears wider as the frequency decreases, measuring 260\,kpc at 1.5\,GHz and 600\,kpc at 144\,MHz. This behavior is a characteristic feature of relics and is attributed to the cooling of electrons in the downstream region of the shock front \citep[e.g.,][]{Markevitch2005,vanWeeren2016a,Rajpurohit2020a}.

In the high-resolution images, the NW relic shows two distinct, bright parallel filaments, which appear broken or fragmented, similar to the Sausage relic \citep{Gennaro2018} and the double threads in the Toothbrush relic \citep{Rajpurohit2018,Rajpurohit2020a}. High-resolution observations often reveal filamentary features in radio relics. These filaments could be tracing the complex shock morphology,  the underlying magnetic field or MHD turbulence \citep{Paola2020a,Wittor2021,Wittor2023}. Alternatively, they can be related to AGN bubbles buoyantly transported outside and ---at some point---crossed and compressed/stretched by a shock. At the southwest edge of the relic, there exists a compact source; however, its morphology does not suggest any connection to the NW relic. 

The low-resolution images have a common resolution of 20\arcsec~ (in Figure\,\ref{high_res} right). R2 is only detected at 144\,MHz, and R1 is detected at all the observed frequencies. Morphologically, R1 and R2 appear disconnected, suggesting that they are very likely two separate structures. 

R3 is situated symmetrically to the northwest relic, suggesting the possibility of it being a double relic system. The LLS of R3 at 144 MHz is 950\,kpc. We determined the size from a high-resolution LOFAR map to prevent potential contamination from the low surface brightness halo emission. In Figure\,\ref{radioxray} left panel, we present the Subaru/Hyper-Suprime Cam \citep{Kim2023} zoom-in view of R3 with 144 MHz radio contours overlaid. There are 10  bright spots within R3 (shown with circles). The absence of optical counterparts, associated with bright spots within R3, is indicated by the dashed cyan circles. The compact nature of the southwest tip of R3 suggests that it might be the core of a tailed-angle galaxy. However, we did not find any optical counterpart. Additionally, the morphology of the radio emission does not resemble that of a tailed radio galaxy. Since an elliptical galaxy in the cluster should be easily detectable at that core, this rules out that R3 is a tailed galaxy.

At 1.5\,GHz, 650\,MHz and 144\,MHz frequencies, there is evidence of diffuse low surface brightness emission that appears to extend in the north-to-south direction, see Figure\,\ref{high_res} right panel for labeling. It has an LLS of 1 Mpc at 1.5 GHz and 1.7\,Mpc at 650\,MHz. We emphasize that above 650\,MHz, R3 is apparently separated from the low surface brightness emission. At 144 MHz, this emission is further extended filling the central region, and is apparently connected to R3.  Based on its location, combined with its low surface brightness, we suggest that it is likely a radio halo. The spectral analysis in Section\,\ref{radioanalysis} rules out the possibility of the halo being the tail of R3. Since the halo emission is visible in high-resolution images, this allows us to decide the rough boundary of the halo and R3 (dashed line in Figure\,\ref{high_res} right-bottom). In our high-frequency images, we also detected at least 40 discrete sources within the central region of the cluster. 


\section{Analysis and discussion}  
\label{analysis}

\begin{figure*}[!thbp]
    \centering
    \includegraphics[width=0.94\textwidth]{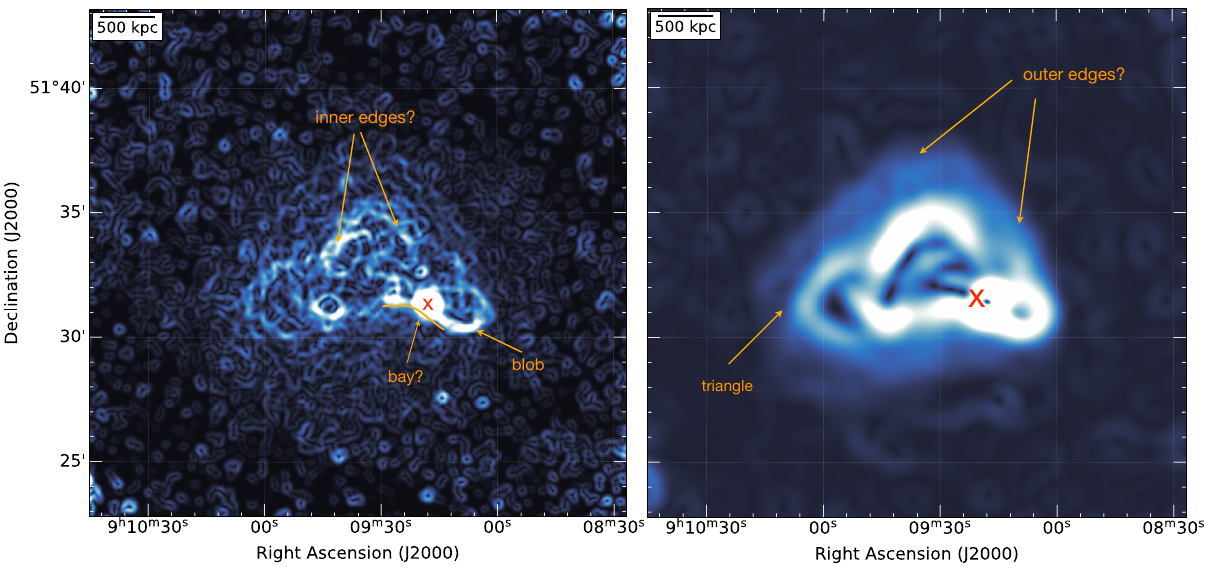}
        \vspace{-0.2cm}
 \caption{GGM filtered \textit{XMM-Newton} images with $\sigma=2$ (left), and $\sigma=12$ (right) pixels. The images reveal the presence of surface brightness edges in the eastern, western, and central regions of the cluster. }
      \label{GGM}
        \end{figure*}  

\begin{figure*}[!thbp]
\centering
\includegraphics[width=0.94\textwidth]{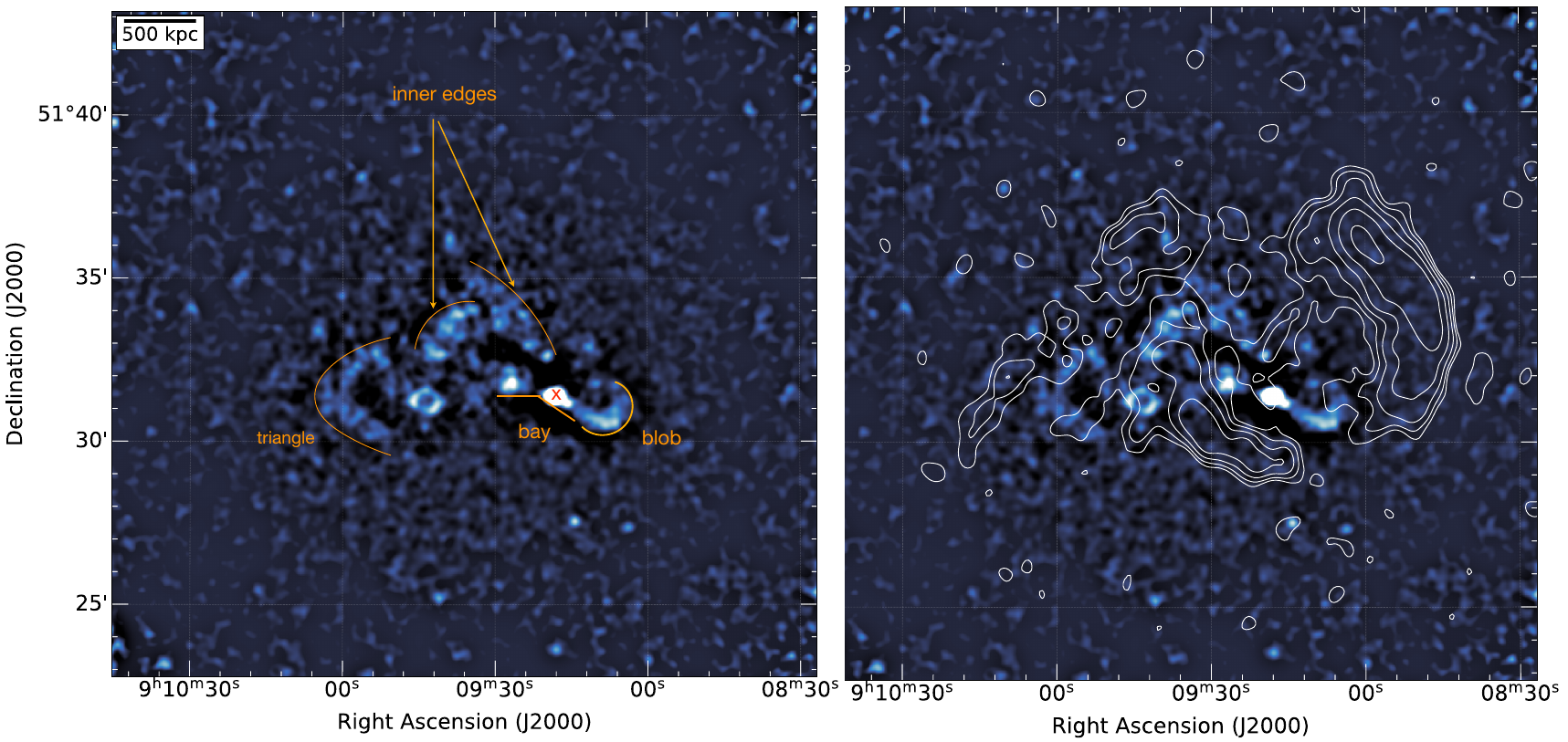}
    \vspace{-0.2cm}
 \caption{Unsharp-masked \textit{XMM-Newton} 0.7-2.0 keV image of the cluster Abell 746 created by subtracting images convolved with Gaussians with $\sigma_1=2\arcsec$ and $\sigma_2=12\arcsec$. The image displays the sharp edges in the X-ray surface brightness image. \textit{XMM-Newton} unsharp image overlaid with 144 MHz LOFAR contours (right). Radio contours are the same as in Figure\,\ref{radioxray} right panel.}
\label{unsharp}
\end{figure*}  

\subsection{X-ray: general properties}

Due to the highly disturbed X-ray morphology of the cluster, the assumptions of equilibrium and spherical symmetry are not valid. As a result, obtaining the cluster's mass using the X-ray properties is not straightforward. Hence, we estimated the mass from the average cluster temperature using the relations from \cite{Lovisari2020}. Our estimated cluster mass $M = 3.0 \pm 0.1 \times 10^{14} M_{\odot}$ was derived considering the temperature within $R_{500}$. The estimated mass differs moderately from the mass obtained by SZ measurements which is  $M_{500}=5.34^{+0.39}_{-0.40}\times10^{14} M_{\odot}$ \citep{Planck2016}. 

The centroid shift ($w$) and concentration parameter ($c$) are among the most robust morphological parameters used to characterize the dynamical state of galaxy clusters (e.g., \citealt{Lovisari2017}). 
The centroid shift measures the standard deviation of the projected separation between the X-ray peak and the centroid of the X-ray emission within different apertures.  Low (high) $w$ values point to a relaxed (disturbed) system.   The concentration parameter measures the ratio between the X-ray surface brightness within two different apertures. A higher concentration parameter implies a more centrally concentrated X-ray emission, while a lower value indicates a more extended distribution. The $w(<R_{500})=0.078$ and $c$=SB($<$0.1$R_{500}$)/SB($<R_{500}$)=0.074 measured for the cluster place the cluster in the bottom right quadrant of the $c-w$ diagram \citep{ Cassano2010}, a region known to be mainly associated with clusters hosting radio halos.

  \begin{figure}[!thbp]
\centering
     \includegraphics[width=0.48\textwidth]{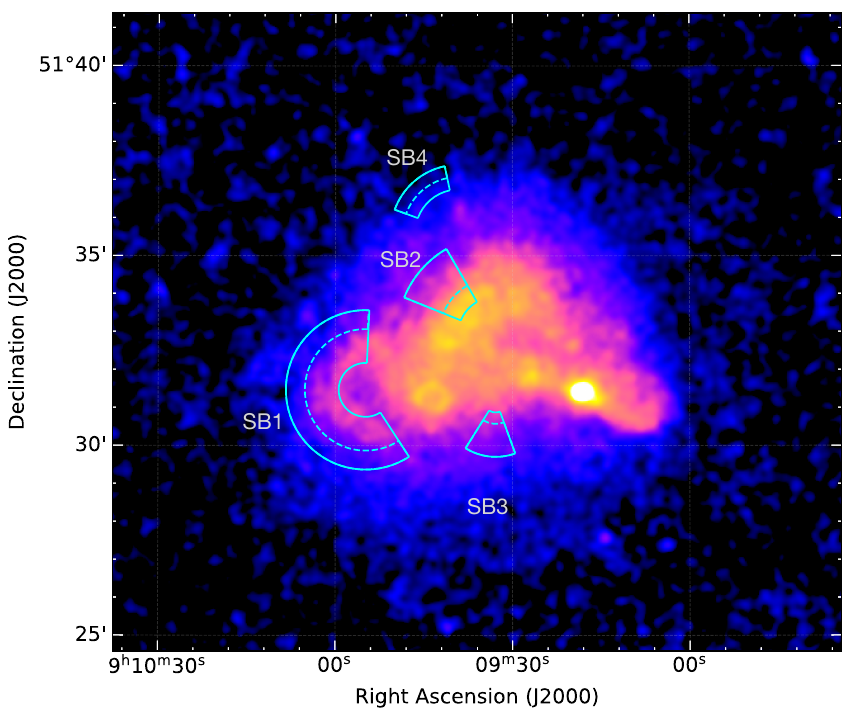}
         \vspace{-0.5cm}
 \caption{\textit{XMM-Newton} image overlaid with four regions of interest are defined and their surface brightness profiles are shown in Figure\,\ref{SB_profiles}. The dashed lines mark the X-ray edges with density jumps.}
      \label{regions}
 \end{figure}

\subsection{Gaussian gradient magnitude filter and unsharp mask images}

Gaussian Gradient Magnitude (GGM) filtering images and unsharp masking serve as powerful tools for identifying regions with surface brightness edges, displaying pronounced and abrupt changes in brightness, or detecting substructures \citep{Sanders2016,Walker2016}. Applying the GGM filter to the $0.7-2.0$ keV band \textit{XMM-Newton} image of Abell 746 reveals the presence of notable surface brightness features in the eastern, western, and central regions, as shown in Figure\,\ref{GGM}. The most prominent structures are characterized by triangular and bay-shaped edges, for labeling see Figure\,\ref{GGM}.  The bay structure extends approximately 500 kpc in length. However, the presence of a bright X-ray point source prevents us from determining whether it forms a continuous edge or not.  

To the southeast of the cluster is a triangle-shaped edge (SB1), clearly seen in the original X-ray image. It is approximately 700\,kpc large in size. The north side of the cluster shows inner edges, see Figure\,\ref{GGM}.  At the north and northwest, where the data quality is poorer, are long linear edges, labeled outer edges.  The northern edge is coincident with the edge of R1. To the southwest of the bay structure, there seems to be a blob-like feature. 

The unsharp-masked image of the cluster generated by subtracting images convolved with $\sigma_1=2\arcsec$ and $\sigma_2=12\arcsec$ Gaussian is shown with overlaid LOFAR 144 MHz contours in Figure\,\ref{unsharp}. The image highlights the presence of the same four distinct features, which were already evident in the GGM-filtered images. The northeast tip of R3 seems to be confined within the positions of the two inner edges. There is an apparent anti-correlation between the bright radio and the X-ray features.

\begin{figure*}[!thbp]
\centering
 \includegraphics[width=0.49\textwidth]{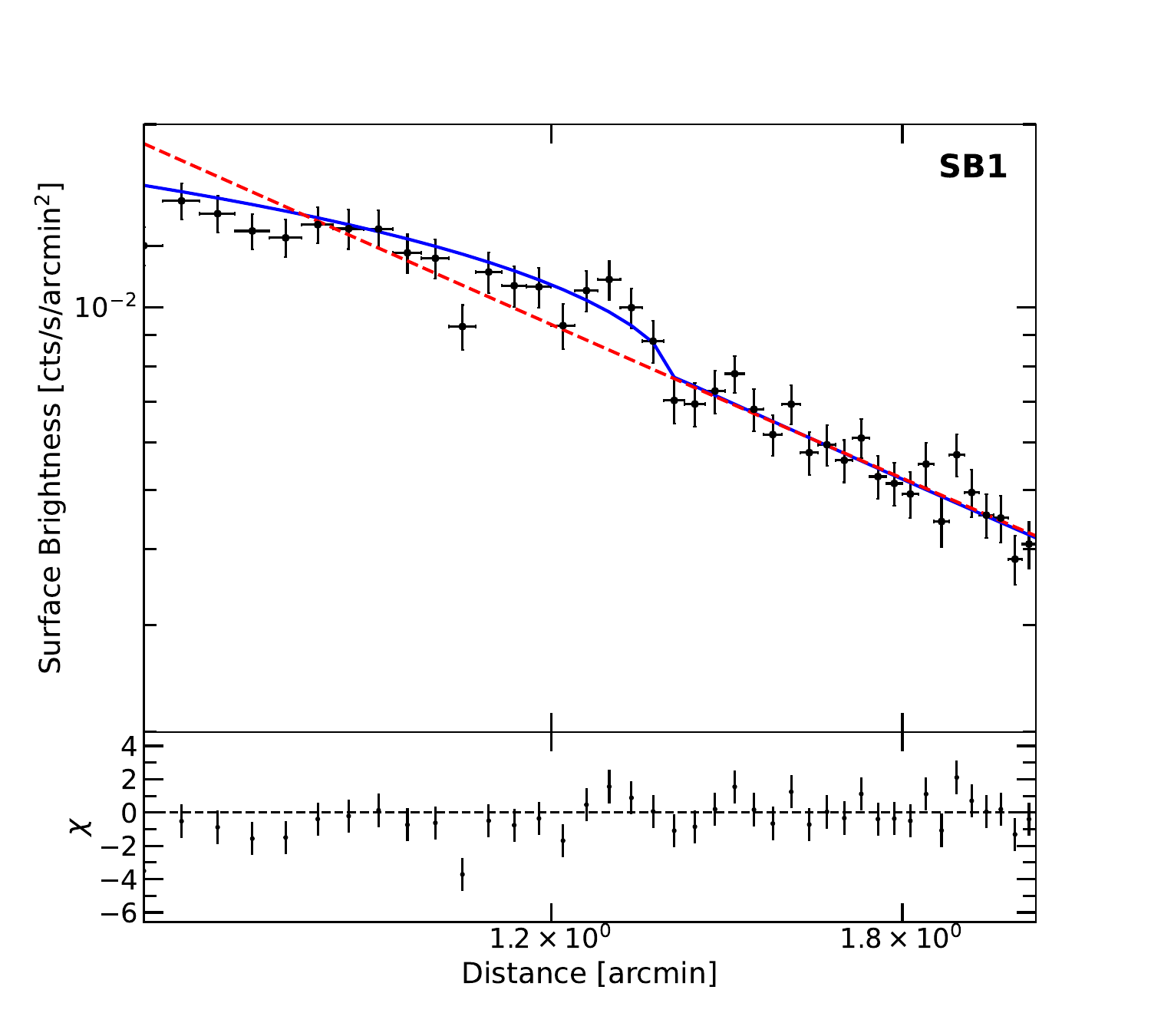}
     \includegraphics[width=0.49\textwidth]{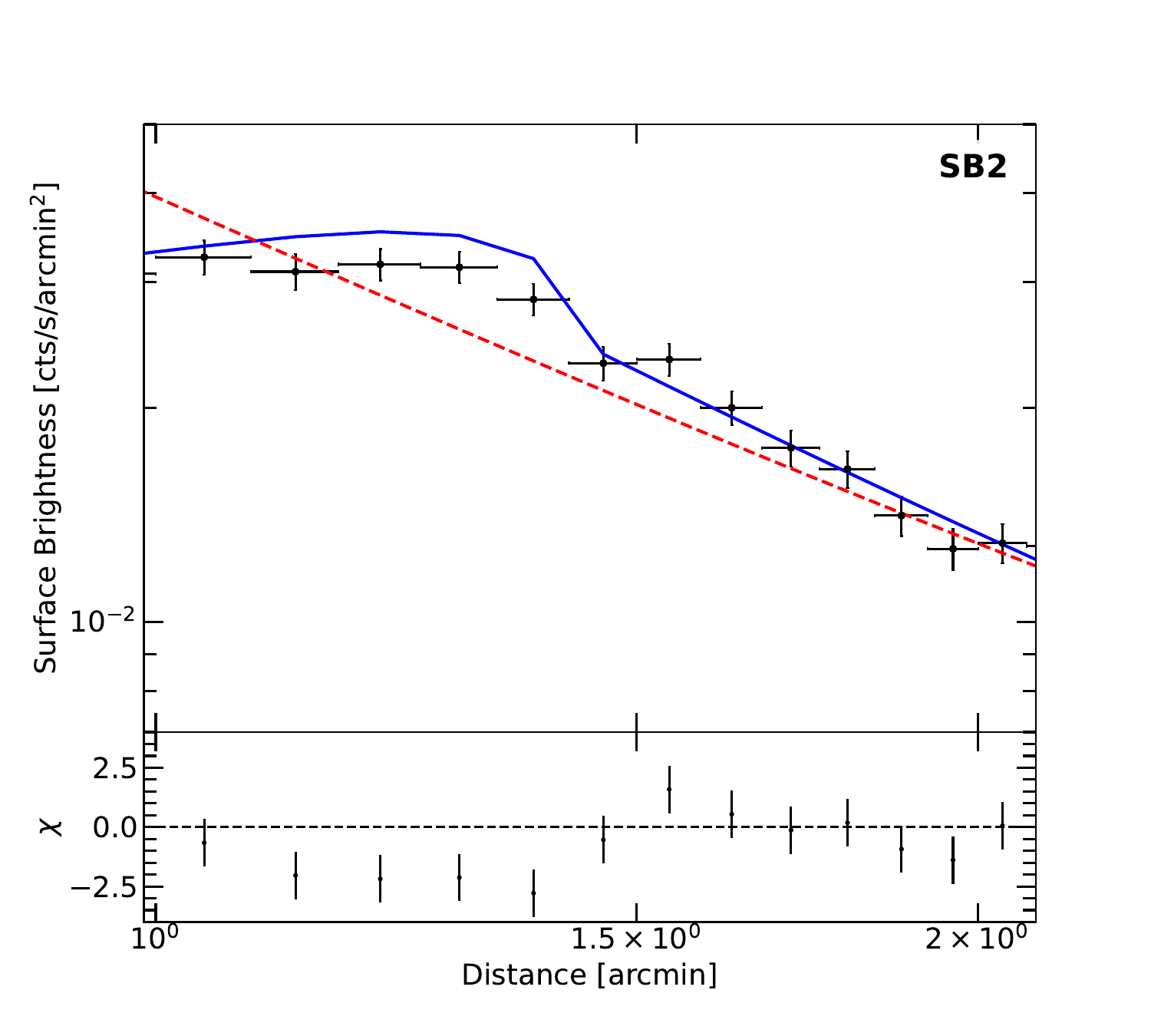}
   \includegraphics[width=0.49\textwidth]{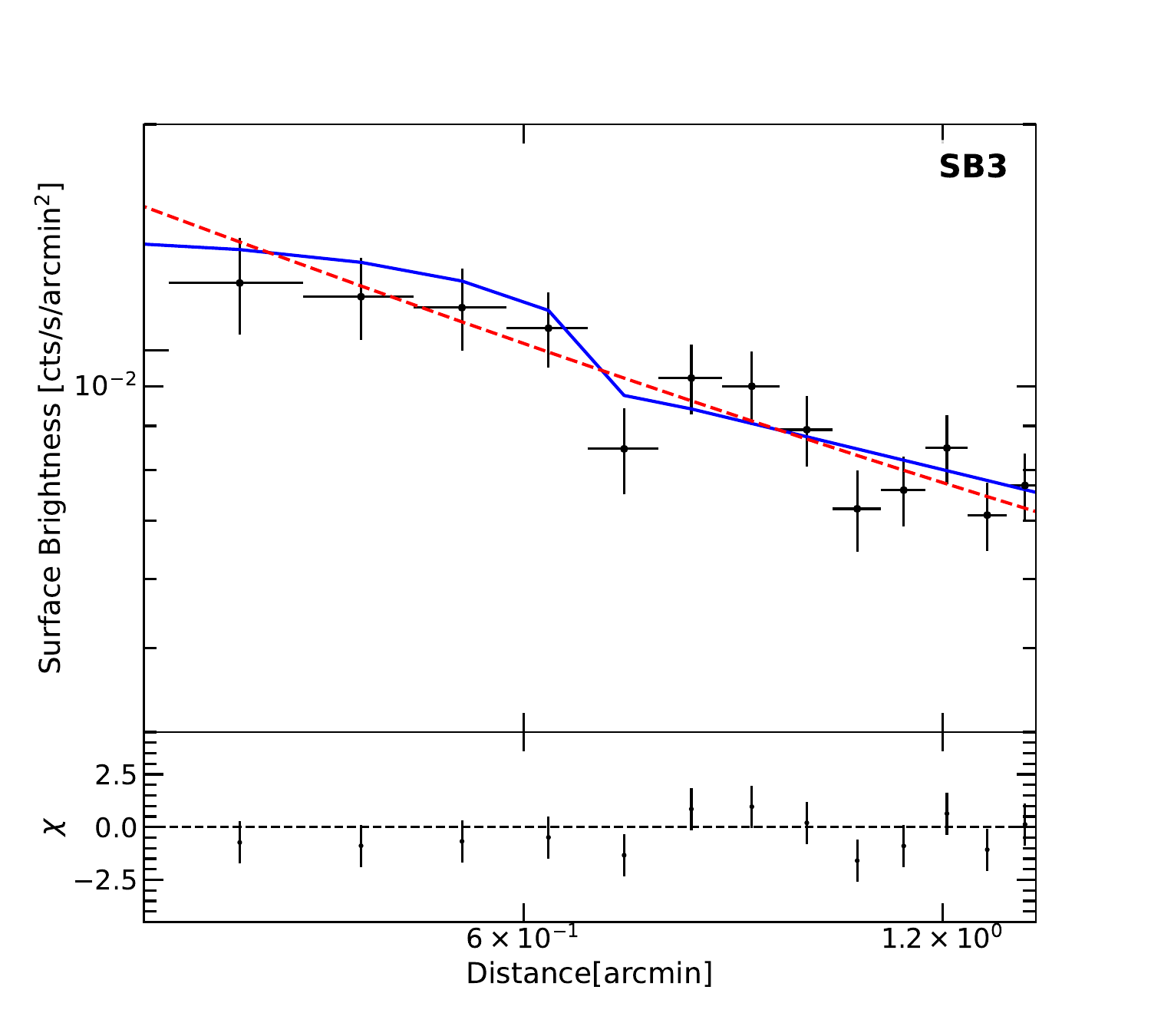}
     \includegraphics[width=0.49\textwidth]{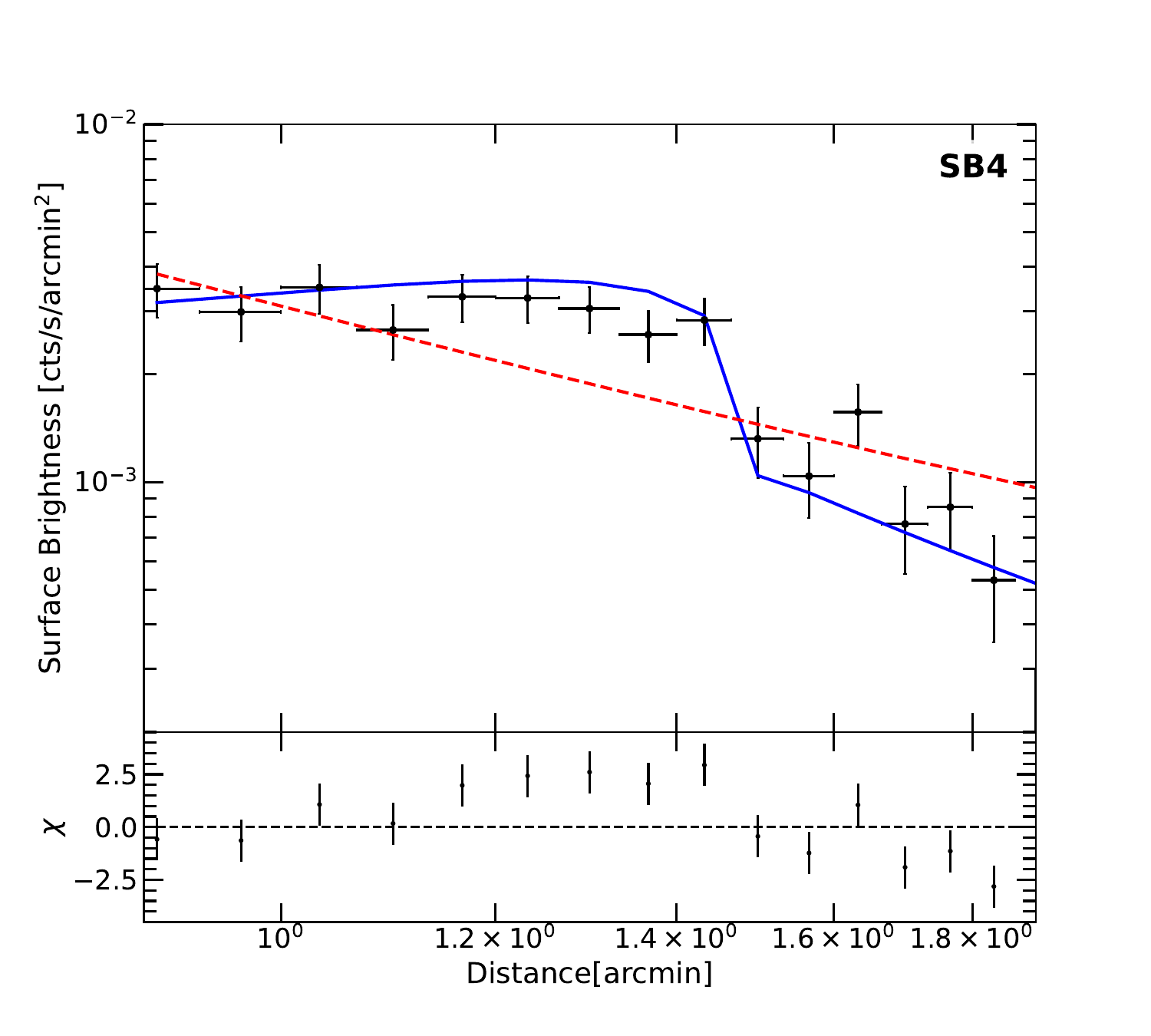}   
         \vspace{-0.5cm}
 \caption{Surface brightness profiles for all four edges (SB1, SB2, SB3 and SB4).  Black data points are from \textit{XMM-Newton}. Blue and red lines are best fitting broken power-law and single power law models, respectively. Residuals on the bottom are from broken power law fits. The sectors where the profiles were fitted and the positions of the relative edges are marked in Figure\,\ref{regions} right panel. }
      \label{SB_profiles}
 \end{figure*}

   \begin{figure*}[!thbp]
    \centering
                \includegraphics[width=1.0\textwidth]{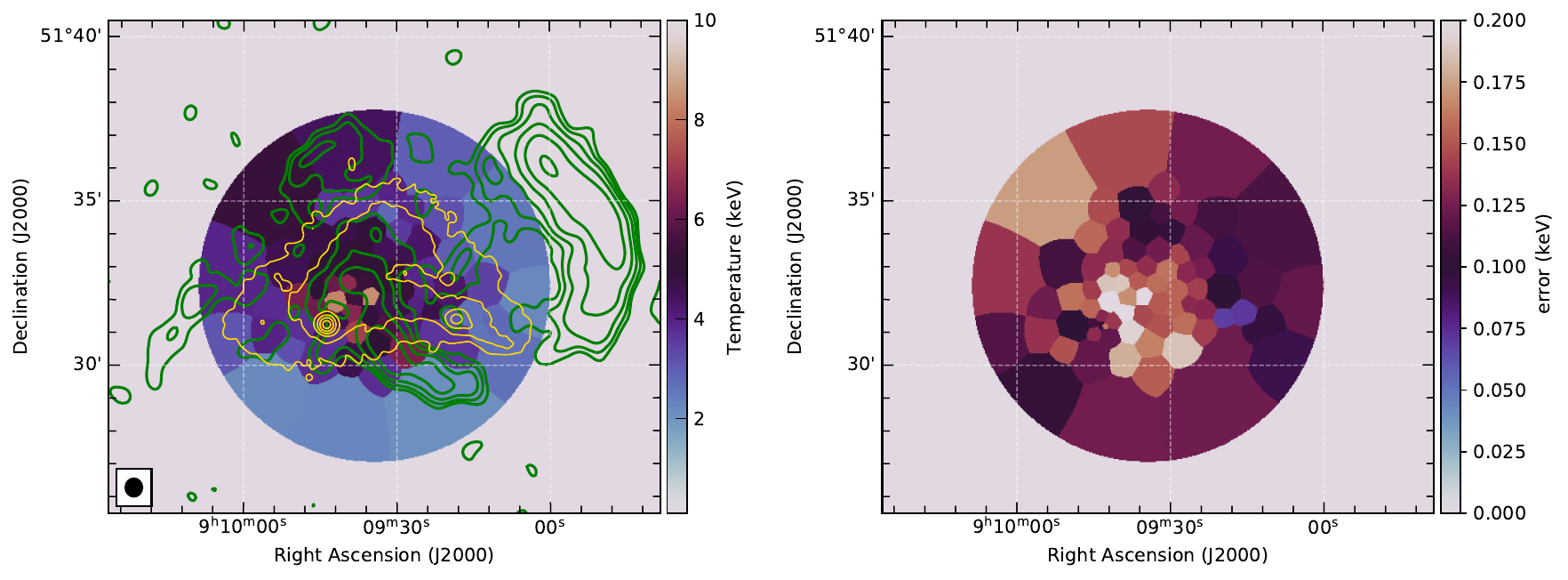}
                    \vspace{-0.5cm}
 \caption{\textit{Left:} Projected temperature map and the relative uncertainties (right), overlaid with the 144 MHz LOFAR contours (in green). The X-ray contours are displayed in orange. The image shows that the temperature distribution is asymmetric across Abell 746: the northern part of the cluster has lower temperatures than the southern part. \textit{Right:} corresponding temperature uncertainty map.}
      \label{temp}
  \end{figure*}

\subsection{Surface brightness profiles}
\label{profiles}

The examination of the GGM-filtered and unsharp mask images already provides initial evidence of potential discontinuities in Abell 746.  To confirm the presence of surface brightness discontinuities, we analyze the surface brightness profiles in sectors around the features seen in GGM-filtered and unsharp mask images and at the location of R1, R2, and R3.  The \textit{XMM-Newton} point-source masked (without dmfilth) exposure corrected image was used to extract surface brightness profiles. We used {\tt Pyproffit} \citep{Eckert2020} to extract and fit the surface brightness profiles. Following the work of \cite{Markevitch2007}, the surface brightness profiles were fitted using broken power-law density models across the edges. The model assumes that the X-ray emissivity is  proportional to the density squared and can be used to describe a density jump linked to a shock or cold front: 

\begin{equation*}
n(r) = \begin{cases}
    \mathcal{C}n_0 \left(\frac{r}{r_i}\right)^{\alpha_2} & \text{if } r \leq r_i ,\\
     n_0 \left(\frac{r}{r_i}\right)^{\alpha_1} & \text{if } r > r_i,
\end{cases}
\end{equation*}
where $\alpha_1$ and $\alpha_2$ define the slopes of the power-laws and the subscripts 1 and 2 denote the upstream and downstream regions, respectively. The electron density jump is characterized by the parameter $\mathcal{C}=n_2/n_1$. The $n_0$ is a normalization factor, $r$ denotes the radius from the center of the sector, and $r_i$ is the position of the jump. 

In the case of a shock, $n_2/n_1$ is related to the Mach number  ($\mathcal{M}$) and can be determined by using the Rankine-Hugoniot  jump condition
\begin{equation}
\mathcal{M_S} =\sqrt{\frac{2C}{\gamma + 1 - C(\gamma - 1)}},
\label{S_M}
\end{equation}
where $\gamma$ is the adiabatic index of the gas and is assumed to be $5/3$ (i.e., a monoatomic gas). Using the canonical shock jump conditions for temperature, the nature of surface brightness discontinuities and $\mathcal{M}$ can be obtained through the following relation

\begin{equation}
\mathcal{M_T} = \sqrt{\frac{\left(8\frac{T_2}{T_1} - 7\right) + \left[\left(8\frac{T_2}{T_1} - 7\right)^2 + 15\right]}{5}},
\label{T_M}
\end{equation}
where $T_1$ and $T2$ represent the upstream and downstream temperatures, respectively. In cases where the ratio $T_1/T_2>1$, this indicates the presence of a shock front. Conversely, when $T_1/T_2<1$, it implies the occurrence of a cold front.

As shown in Figure\,\ref{SB_profiles}, \textit{XMM-Newton} surface brightness profiles reveal the presence of four edges. Figure\,\ref{regions} displays the sectors for which the fitting was performed. The modeling was done assuming a fixed background level of zero because the \textit{XMM-Netwon} background was already subtracted. The best-fitting, double-power-law model reveals the presence of a density jump at four edges, see Table\,\ref{Tabel:fitting}.

For SB1, we find that $\mathcal{C}=1.27\pm0.12$ and a temperature jump from $T_1= 3.18^{+0.45}_{-0.35}$ to $T_2= 3.78^{+0.41}_{-0.37}$. This suggests the presence of a shock front. For SB2 and SB4, we measure $\mathcal{C}=1.48\pm0.28$ and $\mathcal{C}=1.6\pm0.6$, respectively. The low counts do not allow us to derive the pre and post temperatures at SB4. However, the density jump is detected at the northern edge of R1 suggesting a shock with $\mathcal{M}_{S}=2.4\pm0.8$. 

For SB2, the temperature jump from $T_1= 4.16^{+0.49}_{-0.48}$ to $T_2= 5.57^{+0.73}_{-0.58}$. Using Equation\,\ref{T_M}, we obtained  
$\mathcal{M}_{T}= 1.36^{+0.27}_{-0.23}$, respectively. Thus, the results from the temperature and density jump hint that the SB2 edge traces a weak shock front.

\setlength{\tabcolsep}{11pt}
\begin{table*}[!htbp]
\caption{Surface Brightness Profile}
\begin{center}
\begin{tabular}{ l  c  c c c  c c}
  \hline  \hline  
\multirow{1}{*}{}& \multirow{1}{*}{SB1}&  \multirow{1}{*}{SB2} & \multirow{1}{*}{SB3}  &\multirow{1}{*}{SB4}   \\  
\hline
Sector center  Right Ascension &$137^\circ4795$&$137^\circ3877$&$137^\circ3877$ & $137^\circ4137$\\
Sector  center Declination  &$51^\circ5239$&$51^\circ5482$&$51^\circ5195$&$51^\circ5928$\\
Opening angle&$82^\circ-303^\circ $&$120^\circ-158^\circ$&$238^\circ-290^\circ $&$102^\circ-160^\circ$\\ 
Fitting range&0.9\arcmin-2.4\arcmin &0.9\arcmin-1.3\arcmin & 0.33\arcmin-1.39\arcmin&1\arcmin-1.8\arcmin\\ 
 $n_2/n_1$&$1.27\pm0.12$&$1.48\pm0.28$& $1.6\pm0.6$&$2.4\pm0.8$\\
$r_{edge}$ &1.35\arcmin&1.37\arcmin&0.63\arcmin& 1.4\arcmin&\\ 
$\chi^2/d.o.f.$ (broken power law& 53.7/39 &9.9/12&8.7/8&10.7/10&\\ 
$\chi^2/d.o.f.$ (single power law)&32.5/16 &26.6/12&8.9/8&20.7/7 &\\

\hline 
\end{tabular}
\end{center}
\label{Tabel:fitting}
\end{table*} 

For the southwest edge (SB3), we find that the surface brightness jump is not that significant, see Table\,\ref{Tabel:fitting. There is a temperature jump from $T_1= 6.08^{4.44}_{-2.04}$ to $T_2= 3.79^{+1.12}_{-0.92}$. However, the error on the putative pre-shock temperature is too large to confirm or rule out a shock. Hence, we consider it to be a candidate discontinuity}. Moreover, the GGM-filtered and unsharp mask images show other arc-like features, as shown in Figures\,\ref{GGM} and \ref{unsharp}, labeled inner and outer edges, however, we do not find a discontinuity in the SB profile fitting.

\begin{figure*}[!thbp]
    \centering
    \includegraphics[width=0.473\textwidth]{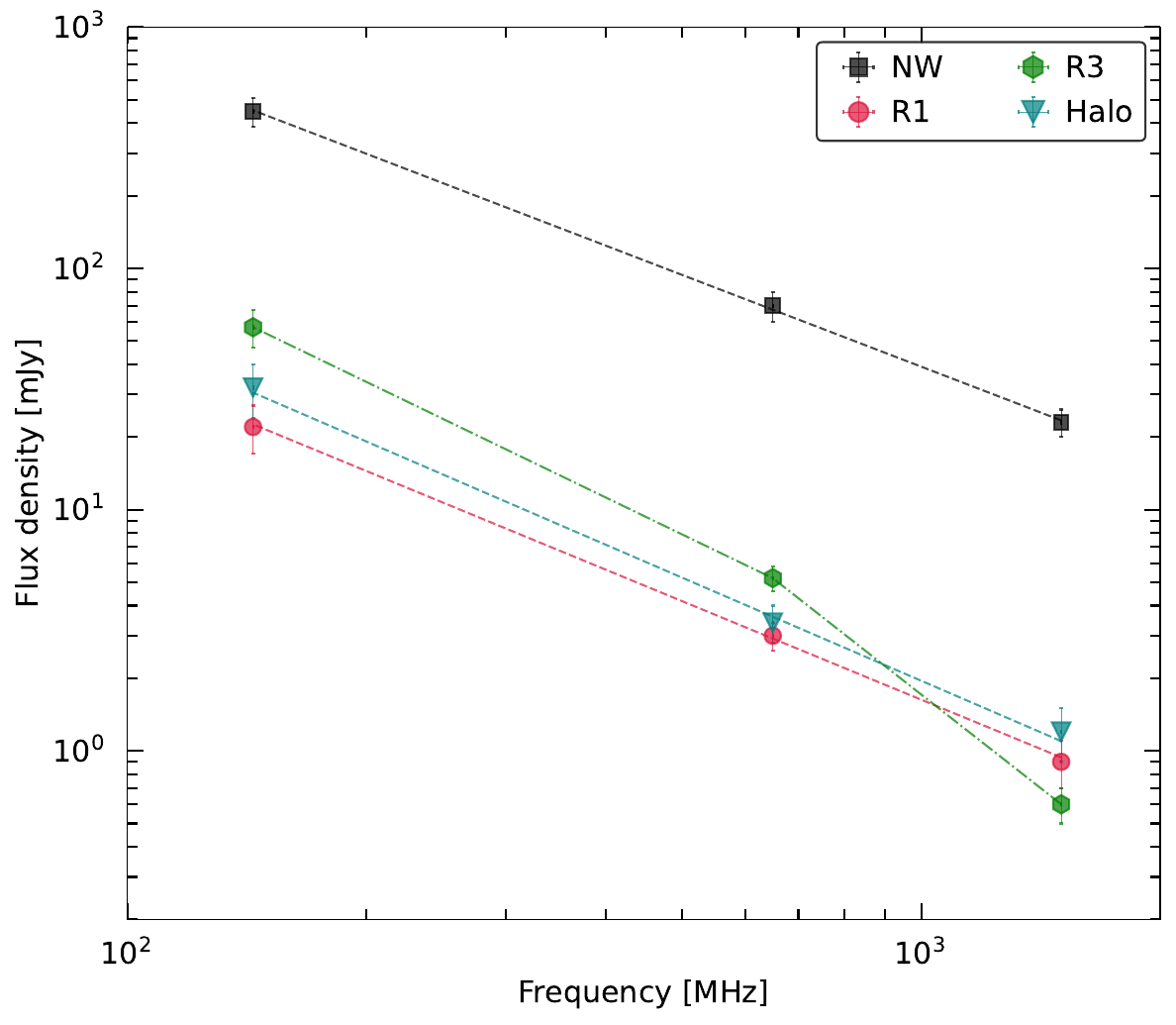}
    \includegraphics[width=0.485\textwidth]{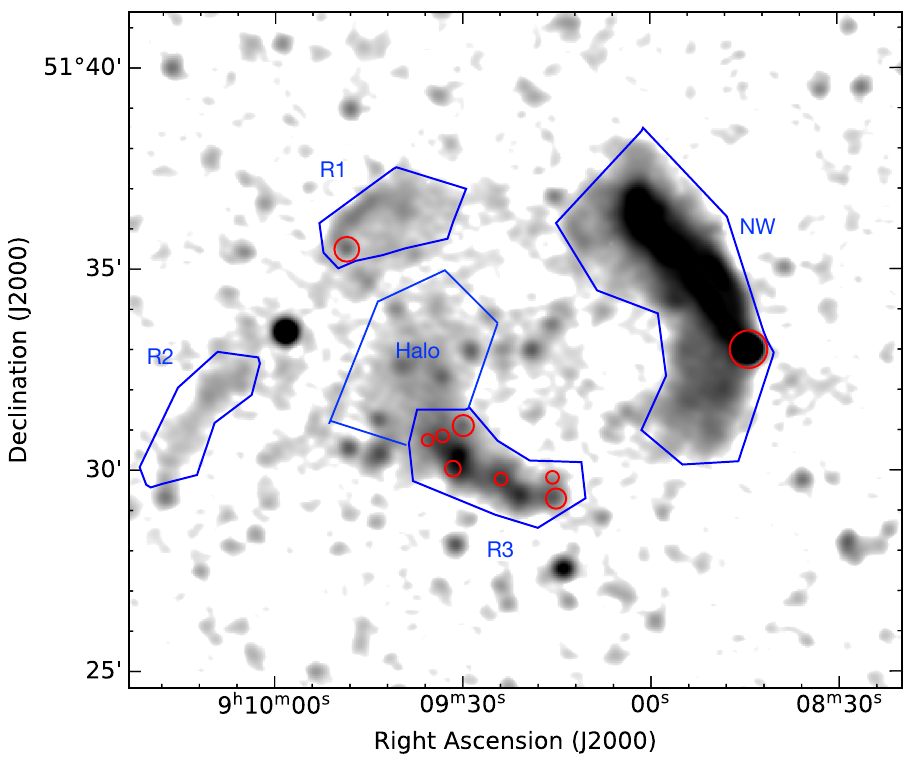}
 \caption{\textit{Left}: Integrated radio spectra of NW, R1, R3, and the halo measured between 144\,MHz and 1.5\,GHz. Dashed lines show the fitted power law and the dotted dash a curved spectrum. The spectra for NW, R1, and the halo are well described by a single power-law. The spectrum of R3 shows a high-frequency steepening and cannot be explained by a single power law.  \textit{Right:} LOFAR 20\arcsec\ resolution image depicting the regions where the integrated flux densities were measured. Flux density contributions from compact sources (shown with red circles) within NW, R1, and R4 were manually subtracted from their total flux densities.}
      \label{spectrum}
  \end{figure*}

\subsection{ICM temperature distribution}
Abell 746 is a very disturbed cluster, therefore a standard radial profile analysis would only provide a partial view of the temperature distribution. We therefore investigated its 2-dimensional spectroscopic properties. To this end, we subdivided the $R_{500}$ volume into small regions from which spectra were extracted. The regions were obtained using the Weighted Voronoi Tessellation (WVT) binning algorithm by \citet{vor06}, which is a generalization of the \citet{2003MNRAS.342..345C} Voronoi binning algorithm, by requiring a signal-to-noise $S/N\sim30$, which allowed us to estimate the temperature with a statistical uncertainty of 10-20\%  \citep[see][for more details]{Lovisari2019}.  

The resulting temperature map, with overlaid X-ray and radio contours, along with the corresponding uncertainties, is shown in Figure~\ref{temp}. The disturbed morphology of the cluster is highlighted by the temperature variation, in particular in the core region (within 2-3\arcmin). Despite the `global' temperature of the cluster being 4\,keV, it is not isothermal at this temperature. We find that the cluster shows an asymmetric temperature distribution; the southern part of the cluster appears to have higher temperatures than the northern region ($\leq4\,\rm  keV$). The temperature at the northern tip of R3 is the highest, reaching $9\rm\, keV$. A possible explanation for this could be a shock front, which leads to the heating of the ICM. The presence of a 0.9\,Mpc elongated diffuse source (i.e., R3) further supports this hypothesis.

The V-shaped region has a temperature of $\rm kT \sim 5\,keV$ which is hotter than its immediate surroundings. To the west of the V-shaped region, the temperature decreases  $\rm kT \sim 3.5\,keV$. There is tentative evidence suggesting that a large region, located southeast of the SB1 edge, exhibits lower temperatures, see Figure~\ref{temp} left panel. The same trends are observed for SB3.

\subsection{Integrated radio spectra}
\label{radioanalysis}

We combined our new uGMRT and VLA observations with published LOFAR HBA data \citep{Botteon2022} to investigate the spectral characteristics of the diffuse radio sources in the field.  We created images with a common inner uv-cut of $100\lambda$ chosen as the minimum well-sampled baseline in the uGMRT Band4 data. This ensures we recover the flux density on the same spatial scales at all observed frequencies. It is a known fact that different resolution and imaging weighting schemes bias the flux density measurements \citep[e.g.,][]{Stroe2016,Rajpurohit2018}.  Therefore, we imaged each data set at a common resolution and with ${\tt robust} =-0.5$.

The radio spectrum has only been measured with $\geq3$  frequencies for a minority of relics and halos \citep{vanWeeren2012a,Stroe2013,Stroe2016,vanWeeren2016a,vanWeeren2017a,Pearce2017,Gennaro2018,Stuardi2019,Loi2020,Rajpurohit2020a,Rajpurohit2021a,Luca2021,Rajpurohit2023,Bonafede2022,Gennaro2023}.  To examine the integrated spectrum of diffuse sources in Abell 746, we consider 20\arcsec\ resolution images for flux density measurements. The same maps are used to create a low-resolution spectral index map.

  \begin{figure*}[!thbp]
    \includegraphics[width=0.49\textwidth]{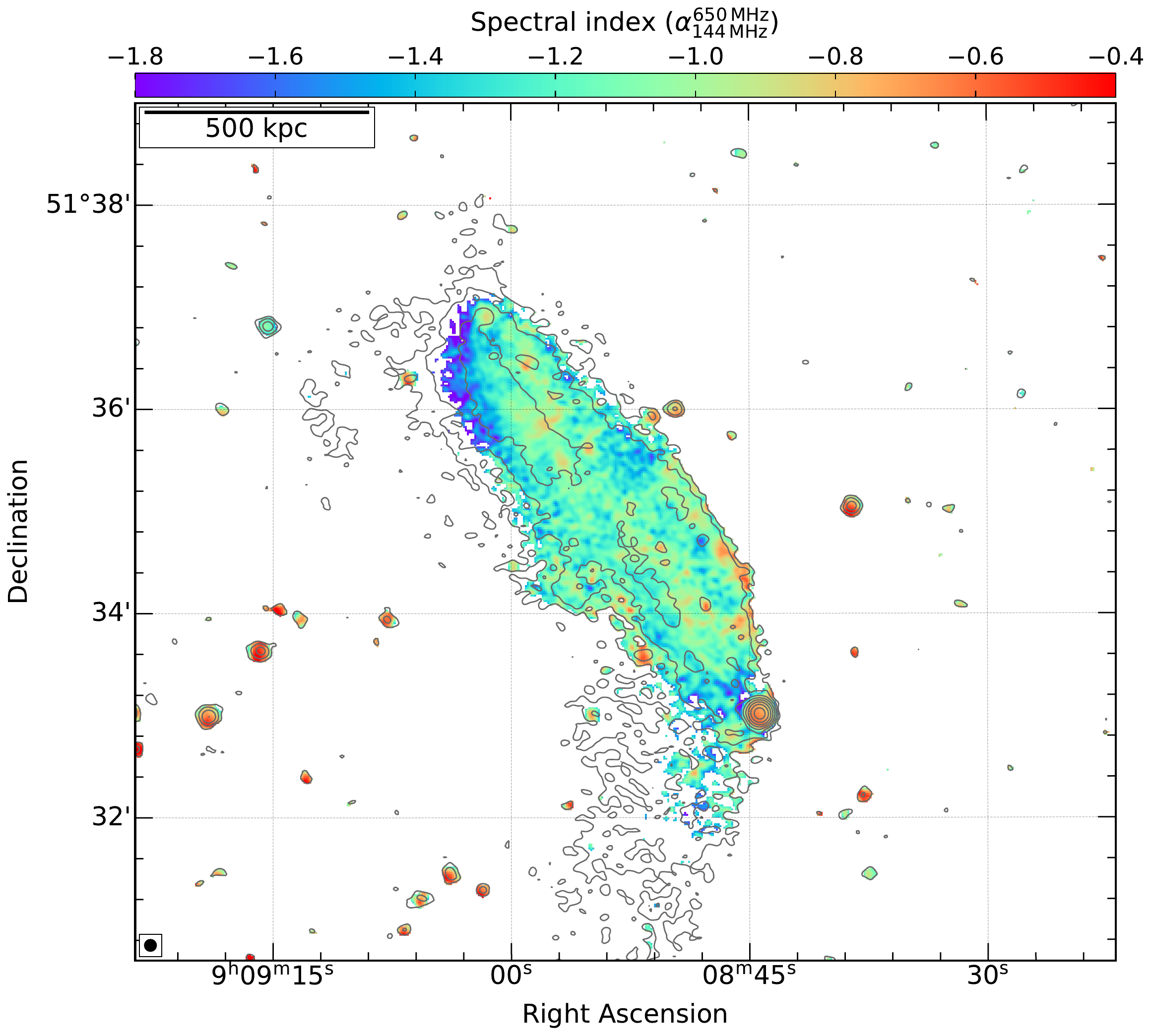}
     \includegraphics[width=0.46\textwidth]{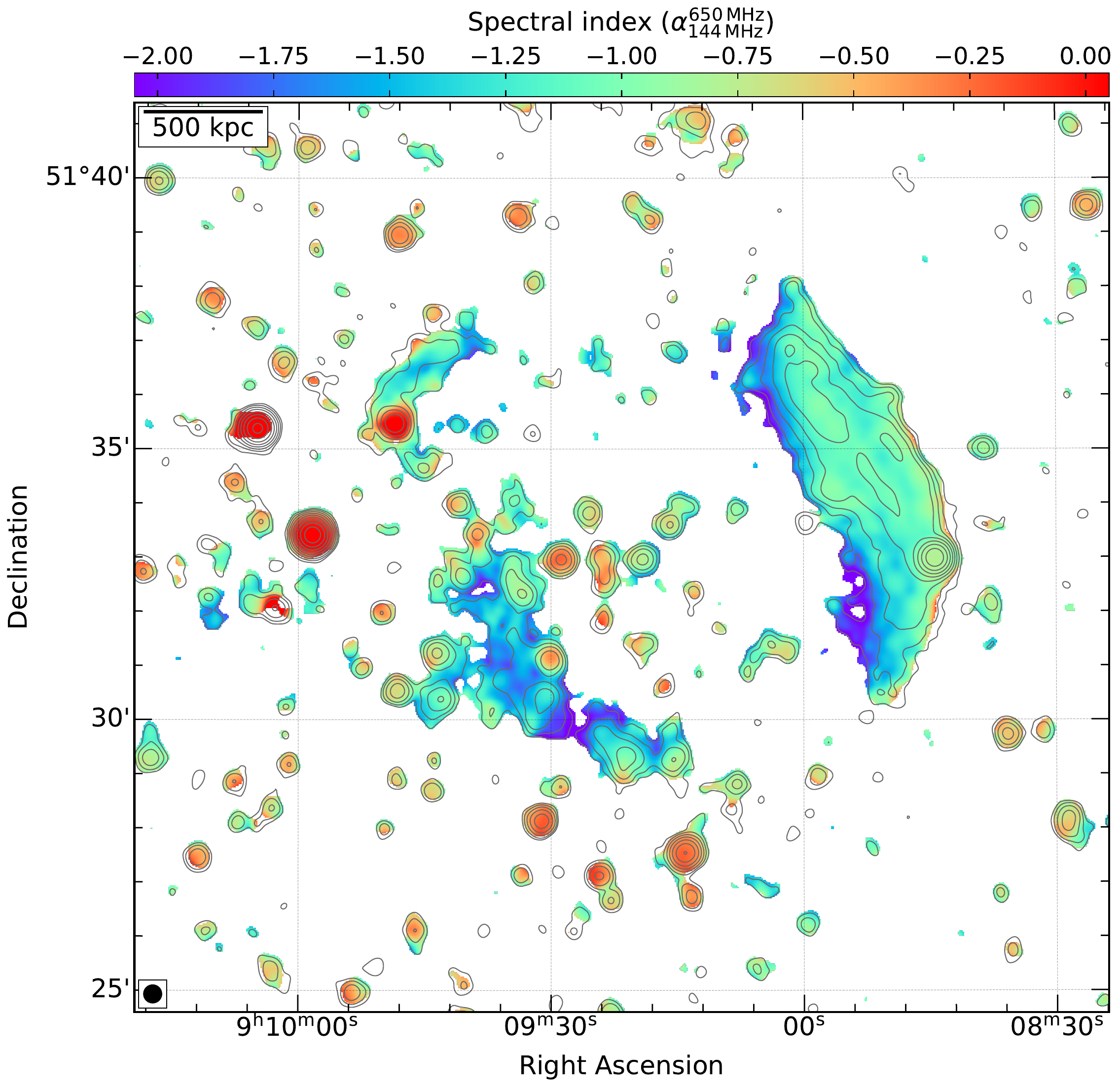}
         \vspace{-0.2cm}
 \caption{\textit{Left:} Spectral index map of the NW relic between 144 MHz and 650 MHz at 7\arcsec resolution.  \textit{Right:} Spectral index map of the Abell 746 at 20\arcsec resolution. Black contours are drawn at levels of  [1, 2, 4, 8, . . .] $\times$ 4.0$\sigma_{{\textrm{\scriptsize rms}}}$  and are from
the LOFAR 144 MHz image. Pixels with values below 3.0$\sigma_{{\textrm{\scriptsize rms}}}$ maps were blanked. The image properties are given in Table\,\ref{imaging}, IM2, IM5, IM4, and IM8.}
      \label{index}
  \end{figure*}  

In Figure\,\ref{spectrum} left, we show the integrated spectra of NW, R1, R3, and the halo. The regions used for flux density extraction are shown in the right panel. The overall spectrum of the NW relic and R1 can be described by a simple power law within the frequency range of 144 MHz to 1.5 GHz. In the stationary shock approximation (i.e., the cooling time of electrons is considerably shorter than the timescale on which the shock strength or geometry changes), the integrated spectrum, $\alpha_{\rm int}$, is steeper than the injection spectrum $\alpha_{\rm inj}$, 

\begin{equation}
\alpha_{\rm int}=\alpha_{\rm inj}-0.5.
\label{stationary}
\end{equation} 
Moreover,  the integrated index is related to the Mach number of the shock as \citep{Blandford1987}: 
\begin{equation}
\mathcal{M}=\sqrt{\frac{\alpha_{\rm int}-1}{\alpha_{\rm int}+1}}.
\label{int_mach}
\end{equation} 
Simulations have consistently shown that the slope of the radio spectrum exhibits non-linear features for non-planar shocks, such as spherically expanding shocks \citep[e.g.,][]{Kang2016a}. This is because these shocks are expanding into a medium with decreasing density and temperature. In multifrequency radio observations, the majority of relics exhibit a power law behavior across a wide frequency range and adhere to the stationary shock condition \citep{Rajpurohit2020a,Rajpurohit2020b,Loi2020,Rajpurohit2021a}. The integrated spectral index of the NW relic is $-1.26\pm0.04$ and for R2 $-1.35\pm0.07$. Like other well-known relics, sources NW and R1  in Abell 746 follow the stationary state shock condition. Using Equation\,\ref{int_mach}, we obtained a Mach number of $\mathcal{M}_{NW}=2.9^{+0.3}_{-0.1}$ and $\mathcal{M}_{R1}=2.6\pm0.2$. Our \textit{XMM-Newton} data indicates the presence of a discontinuity that aligns with R1. The X-ray Mach number obtained from the density jump is $\mathcal{M}_{R1,X-ray}=2.4\pm0.8$, which agrees within uncertainties with the radio determined Mach number.

To extract the flux density of R3, we subtracted the contribution from compact sources marked in Figure\,\ref{spectrum} right panel. Additionally, we subtracted the flux density of three compact regions within R3 (marked with cyan circles in Figure\,\ref{radioxray}) that do not show any optical counterparts. The integrated spectrum of R3 exhibits a high-frequency steepening, rather than a power law. Its low-frequency spectral index is $-1.6\pm0.1$, while the high-frequency one is $-2.5\pm0.2$.  It is possible that R3 is observed in projection with the halo emission, resulting in its contribution to the total flux density of R3, particularly at 144 MHz. To minimize the contamination from the halo, we measured the flux density of R3 using a $7\arcsec$ resolution image. This higher resolution allows us to minimize flux contamination from the halo to the greatest extent possible. However, the resulting spectrum of R3 is inconsistent with a single power-law. 

To measure the flux density of the halo accurately, it is necessary to subtract the flux density contribution of the point sources embedded within it. Due to the low flux density of the halo in Abell 746 and the presence of a large number of point sources, modeling and subtracting their contributions from the data is challenging. Furthermore, at 144 MHz, the majority of unrelated sources within the halo are not distinguishable from the surrounding diffuse emission.  Therefore, we used high-frequency images to mark their locations and subsequently manually measured and subtracted their flux density from the halo emission. The resulting values are reported in Table\,\ref{Tabel:Tabel2}. 

The halo's overall spectrum can be characterized by a single power law between 144 MHz and 1.5 GHz, with a slope of $-1.48\pm0.10$. The spectral index of the halo is relatively flat compared to R3, supporting the claim that the radio emission in this region is not associated with R3. 

Using optical, radio, and X-ray observations, \cite{Kim2023} proposed a possible merging scenario that involves two successive mergers with three subclusters, which can explain the presence of three relics. According to their scenario, a near head-on collision along the east-west direction between two subclusters results in the formation of relics R1 and R3. Subsequently, a third subcluster moves from south to north, influencing the trajectories of the first two subclusters and leading to the formation of the NW relic.

Using the flux densities measured at 1.5\,GHz and 144\,MHz, we estimate the total radio power of the halo. The total rest-frame radio powers of the halo are  $P_{\rm 1.4\,GHz}\sim 2\times10^{23} \,\rm W\,Hz^{-1}$ and  $P_{\rm 150\,MHz}\sim 5\times10^{24}\,\rm W\,Hz^{-1}$. The radio power of halos (at 1.4 GHz and 150 MHz) exhibits a well-defined correlation with the mass and X-ray luminosity of the host cluster \citep{Cassano2013,vanWeeren2020,Duchesne2021,Cuciti2021}. The halo in Abell 746 fits well in the radio power versus mass relation at 1.4\,GHz and 150\,MHz.  

\subsection{Spectral index maps}
Spectral index studies of radio relics provide important information on their formation and connection to cluster merger processes. Merger-shock models predict an increasing spectral gradient in the post-shock areas of relics, as seen, for example, in the Sausage and Toothbrush relics \citep{vanWeeren2010,Stroe2013,Hoang2017,Gennaro2018,vanWeeren2011,vanWeeren2016a,Rajpurohit2018,Rajpurohit2021a}. We created high and medium-resolution spectral index maps. For spectral index maps, we only considered a common region where pixels with flux density are above $3\sigma_{\rm rms}$ in both maps.

Figure\,\ref{index} displays the high-resolution spectral index map of the NW relic and the cluster created between 144\,MHz and 650\,MHz, where the resolution is $7\arcsec$ (left) and 20\arcsec\ (right). In the high-resolution spectral index map, specific features show noteworthy characteristics within the NW relic. The northern filament edge displays a relatively flat spectral index ($-0.80$ to $-0.95$), while downstream (i.e., toward the cluster center), the spectrum steepens ($-1.8$).  At the southern filament, however, the spectral index once again becomes flatter, followed by a subsequent steepening. A similar type of trend is seen across the ``double strands" in the Toothbrush relic which may hint at a new injection \citep{Rajpurohit2018,Rajpurohit2020a}. At the eastern tip of the relic, the spectral index is as steep as $-1.8$. The low-resolution spectral index map of displays a clear spectral index gradient toward the cluster center, as expected due to electron aging, consistent with observations of other relics. We emphasize that the low surface brightness regions of the NW relic are recovered in the low-resolution images, therefore the 20\arcsec\ resolution spectral index map shows a clear spectral index gradient.

\begin{figure}[!thbp]
\centering
   \includegraphics[width=0.47\textwidth]{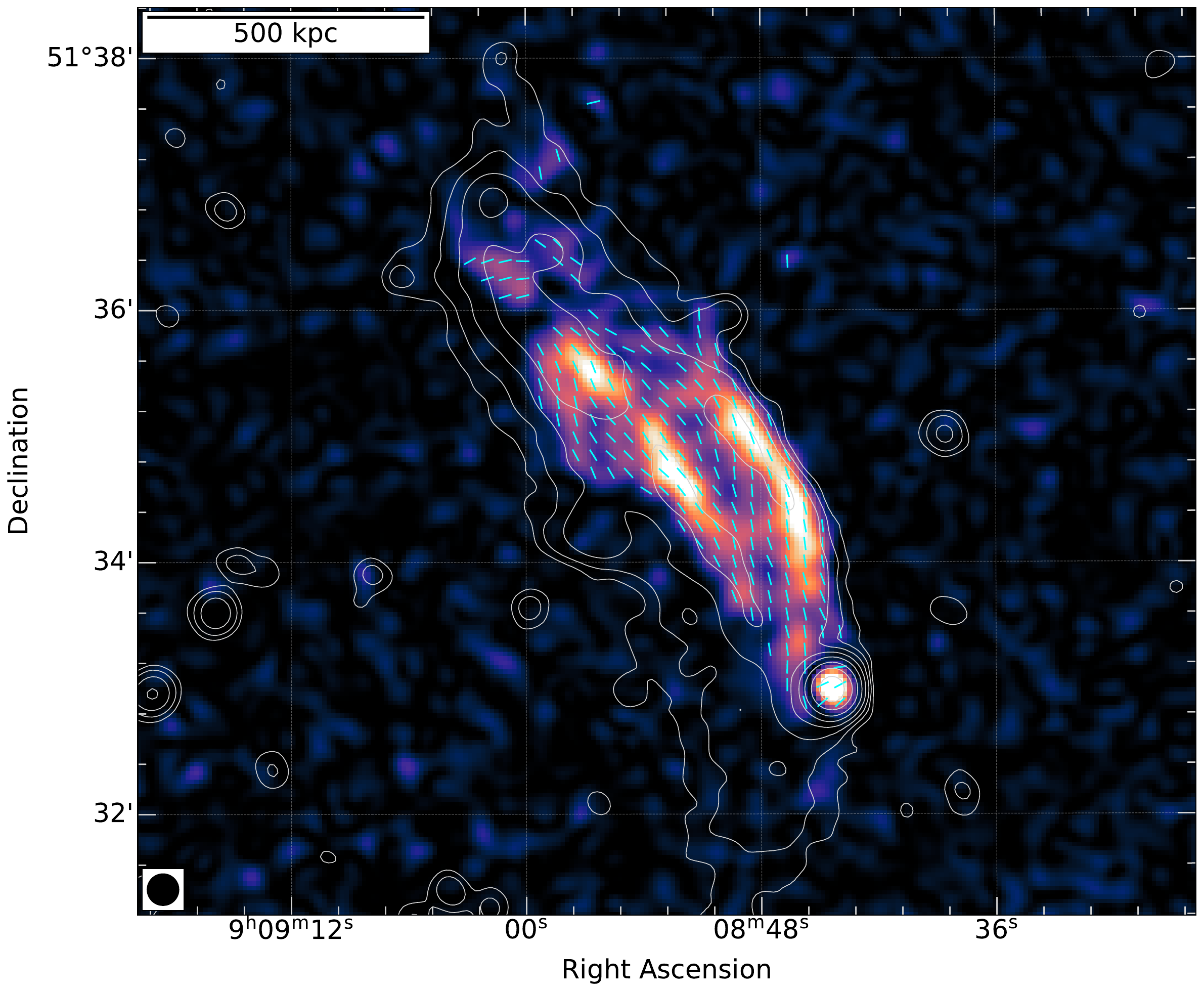}
       \vspace{-0.5cm}
 \caption{Vector map of the NW relic  (at 15\arcsec\ resolution) showing the orientation of the magnetic field vectors overlaid on the linear polarized intensity image. The vectors are corrected for the Faraday rotation due to the Galactic foreground. The Stokes I contours are from the GMRT 15\arcsec\ uv-tapered image and are plotted at levels of  [1, 2, 4, 8, . . .] $\times$ 4.0$\sigma_{{\textrm{\scriptsize rms}}}$ where $\rm {\textrm{\scriptsize rms}}=12\mu\,Jy beam^{-1}$. }  
\label{pol}
\end{figure} 

\begin{figure*}[!thbp]
\centering
          \includegraphics[width=0.95\textwidth]{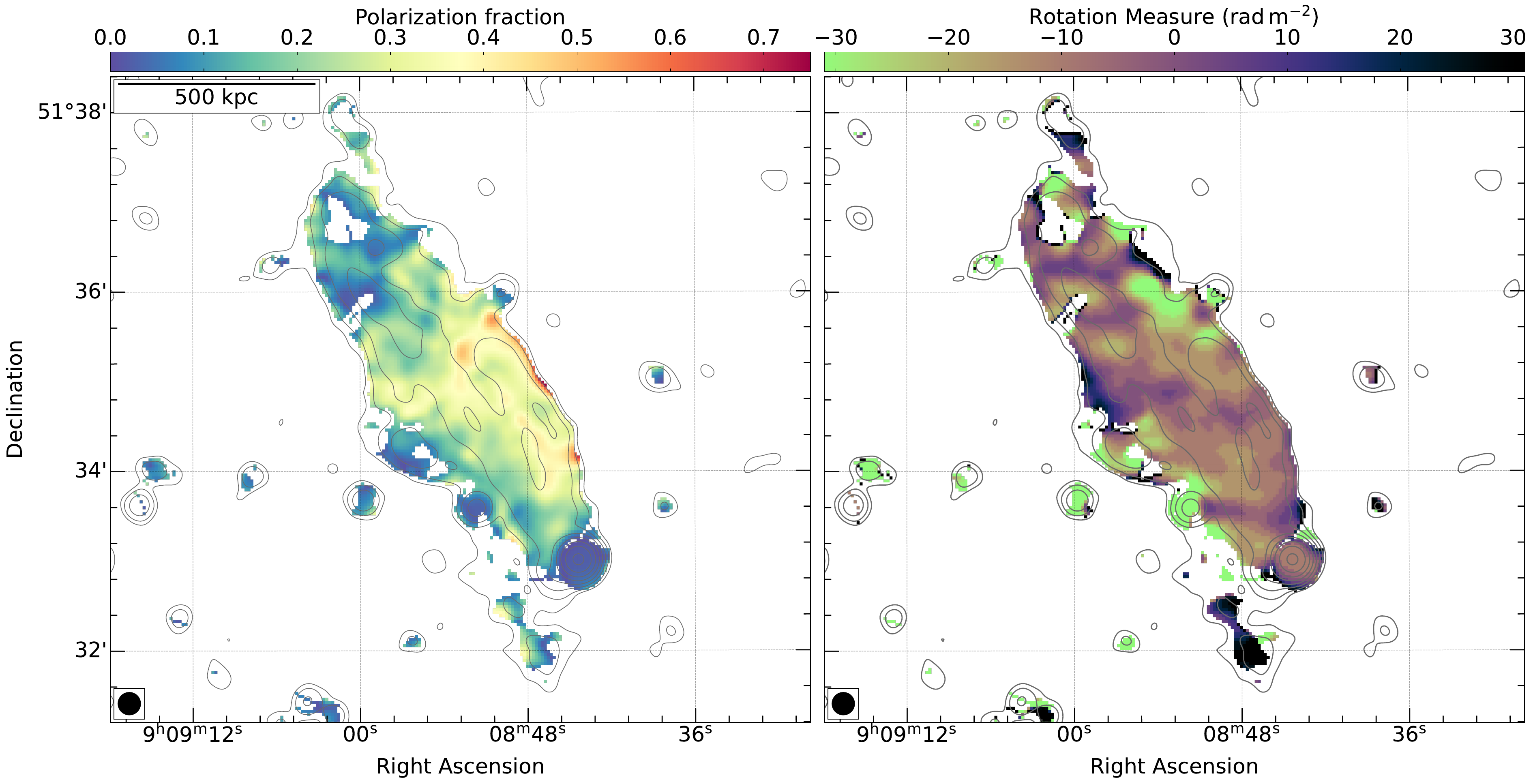}
          \caption{\textit{Left}: The fractional polarization map of the NW relic, obtained at a resolution of 20\arcsec, reveals a high degree of polarization, reaching up to 65\%. The degree of polarization decreases in the downstream regions of the relic. \textit{Right}: Rotation Measure map of the NW relic measured over 1.0-2.0\,GHz using RM-synthesis technique. The RM varies across the relic with coherence lengths of 70-100 kpc.}  
\label{RM}
\end{figure*}

As shown in Figure\,\ref{index} right panel, R1 exhibits a hint of a spectral index gradient towards the cluster center.  The spectral index behavior at R3 is inconsistent with that of a tailed galaxy, where there is typically a spectral steepening in the tail as a function of distance from the AGN due to the aging of CRe. At the southwestern part of R3, there is tentative evidence of a spectral index gradient toward the cluster center. However, the relatively low flux density of R3 at 650 MHz hinders its clear appearance in the spectral index map.

\subsection{Radio power vs LLS}

Radio relic powers (at 1.4\,GHz and 150\,MHz) correlate with LLS, consistent with larger shock surfaces at cluster peripheries \citep{vanWeeren2009,Bonafede2012,deGasperin2014,Jones2023}. The integrated spectral indices of the NW relic and R1 are  $-1.26\pm0.04$ and $-1.36\pm0.07$. This results in the monochromatic radio powers of $3.9\times10^{24}$ and $1.6\times10^{23}$, respectively  at 1.4\,GHz. The radio powers at 150\,MHz are $P_{150 \rm\,MHz, NW}=7.7\times10^{25}$ and $P_{150 \rm \,MHz, R1}=3.8\times10^{24}$.   The NW relic and R1 fall within the known relic relation between the radio power of a relic and LLS. 

The integrated spectral index of R3 between 144\,MHz and 1.5\,GHz is curved.  By adopting the low frequency spectral index of $-1.6$, the estimated radio power of the source at 150\,MHz and 1.5\,GHz is $1\times10^{25}$ and $1\times10^{23}$, respectively. However, this is potentially an underestimate because we exclude the flux density from three compact bright regions (lacking optical counterparts) that are embedded within R3. It fits well with the known correlation between the radio power of a relic and LLS. 


\section{L-band polarization}

\cite{vanWeeren2011} reported that the NW relic in Abell 746 is polarized at 1.4\,GHz and shows aligned magnetic field vectors along its major axis. We present a Faraday Rotation Measure (RM) analysis covering the frequency range of 1-2\,GHz.

We create VLA L-band Stokes IQU cubes at 20\arcsec\ resolution. The same images are used to perform RM-synthesis. These images were obtained employing robust=0.5 weighting and a uv taper of 15\arcsec. As the output images have slightly different resolutions, all images were smoothed to a common resolution, namely 20\arcsec\ resolution. All maps were corrected for the primary beam attenuation. Images with low sensitivity or significant artifact contamination were excluded from the analysis.

The polarization fraction (f), linear polarized intensity (p), and polarization angle ($\chi$)can be determined from the I, Q, and U images via
\begin{subequations}
\begin{align}
p = \sqrt{Q^2 + U^2},\\
f = \frac{\sqrt{Q^2 + U^2}}{I},\\
\chi = 0.5 \tan^{-1}\left(\frac{U}{Q}\right).
\end{align}
\end{subequations}

The resulting polarization intensity map of the NW relic is shown in Figure\,\ref{pol}. At 1-2 GHz,  the relic is polarized over its entire length. The morphology of the polarized emission is similar to the total power emission, in particular, we can see two filaments clearly. No significant polarization is detected for R1 and R3. 

Figure\,\ref{RM} left panel shows that the polarization fraction across the NW relic varies between 1 and 65\%, where it can be measured. The northern filament edge exhibits the highest degree of polarization; this is the region where we anticipate that acceleration is happening. Subsequently, a decrease in the fractional polarization is observed as we move away from this region. At the southern filament, again, the fractional polarization increases. The fractional polarization in the eastern part of the relic is lower compared to the western part. Specifically, at the eastern tip of the relic, the fractional polarization is below 15\%. There is a hint that the degree of polarization decreases in the downstream regions. This is consistent with what has been observed in other relics,  for example, the Sausage, Abell 2744 and MACS\,J0717.5+3745 relics \citep{DiGennaro2021,Rajpurohit2021c,Rajpurohit2022a}. According to simulations, the presence of shock and downstream turbulence is likely to produce the observed trends in polarization \citep{Paola2021,Paola2021b}. As previously reported by \cite{vanWeeren2011}, the magnetic field vectors exhibit a remarkable alignment with the major axis of the relic emission.

RM-synthesis \citep{Brentjens2005} was performed on Stokes IQU cubes with 51 spectral channels using ``{\tt pyrmsynth}".  The RM cube synthesizes a range of RM from $-300\,\rm rad\,m^{-2}$ to $300\,\rm rad\,m^{-2}$ with a bin size of $5\,\rm rad\,m^{-2}$. In Figure\,\ref{RM} (right panel), we show the Faraday map of the NW relic where the RM values vary between $-30$ to $+30$ $\rm rad\,m^{-2}$. The RM across the entire relic shows a well-defined single peak. Additionally, there are some observable small-scale fluctuations in the RM. 
The mean RM across the NW relic is $-10\,\rm rad\,m^{-2}$. At the position of Abell 746, the average Galactic RM contribution is $-20\,\rm rad\,m^{-2}$. Therefore, the detected RMs seem to be consistent with the average Galactic foreground.

\subsection{Magnetic field estimates of the NW relic}

The magnetic field at the NW relic can also be estimated, at least to a zeroth order level, assuming equipartition between the energy densities of cosmic rays in the plasma, and the magnetic energy $\epsilon_{CR}=\epsilon_B$ \citep[e.g.][]{1997A&A...325..898B, 2005AN....326..414B}, which also gives a value close to the minimum of the combined energy densities, $\epsilon_B+\epsilon_{CR}$.  In radio relics it is realistic to assume that the spectrum of cosmic ray electrons in the downstream results from the combination of injection, transport and energy losses.
Following the formalism in \cite{Locatelli2020}, the equipartition estimate of the magnetic field can in this case be obtained by assuming that the magnetic field has the same energy density of the cosmic rays downstream:
\begin{equation}
    \begin{split}
\epsilon_{CR} =    \frac{1}{2}\rho_u\, \frac{\rm v^3_u}{\rm v_d}\, \xi_e\, (1+k) = \frac{B^2}{8\pi} =  \epsilon_B
    \end{split}
\end{equation}
where $k$ is the ratio of energies budget between cosmic ray protons and electrons,  $\rho$ and $\rm v$ are the gas density and shock velocity computed  upstream (${}_u$) and downstream (${}_d$) of the shock front. Based on the total radio spectrum of the relic, we use a shock Mach number $\mathcal{M}=2.6$ to estimate $\rho$ and $\rm v$ at both sides of the shock discontinuity, assuming Rankine-Hugoniot jump conditions.
At NW, $T_d \sim 5.57 ~\rm keV$, which implies $T_u=1.89 ~\rm keV$, and an upstream sound speed of $c_s=709 \rm ~km/s$, and hence a shock speed $v_s=M c_s=1845 ~\rm km/s=v_u$. The downstream gas velocity is thus $v_d=v_u \cdot (M^2+3)/(4M^2)$. If we assume somewhat canonical (albeit far from being certain) values of $\xi_e=10^{-4}$ for the electron injection efficiency at this Mach number,  we get $B_{\rm equip} \approx 0.9 \mu G$ for $k=100$, $B_{\rm equip} \approx 0.2 \mu G$ for $k=10$ and $B_{\rm equip} \approx 0.1 \mu G$ for $k=1$. Manifestly, the various uncertainties on the amount of electrons and protons present in the relic region make this estimate uncertain, and with the additional uncertainties on the estimate of the shock Mach number, a more accurate estimate will be difficult to obtain.

Under a few simplifying assumptions, the magnetic field strength can also be estimated using the observed Faraday dispersion ($\sigma_{\phi}$, \cite{Sokoloff1998,Kierdorf2016}:
 \begin{equation}
  \sigma_{\phi}
  =
  \sqrt{(1/3)}\,0.81\,\langle n_{\rm e}\rangle\,B_{\rm turb}\,(L\,t/f)^{0.5},
\label{sigmaRM}
\end{equation}
 where $\langle n_{\rm e}\rangle$ is the average thermal electron density of the ionized gas in $\rm cm^{-3}$, $B_{\rm turb}$ is the magnetic field strength in $\upmu$G. $L$ and $t$ are the path length through the thermal gas and turbulence scale, respectively, in pc and $f$ the volume filling factor of the Faraday-rotating plasma. For the NW relic, the average $\sigma_{\phi}=10 ~\rm rad\,m^{-2}$. We assume that the thermal density along the line of sight is $10^{-4} ~\rm cm^{-3}$, $L\sim1\,\rm Mpc$, $t=50\,\rm kpc$ and $f=0.5$ \cite{Murgia2004,Kierdorf2016}. By inserting all values in Equation\,\ref{sigmaRM}, we obtained $B=0.7 \mu \rm G$. This is consistent with the estimates from the equipartition using $k=100$.


\section{Origin of NW, R1, R2, and R3}
\label{origin}

Based on the morphology of the diffuse source NW, its location, high degree of polarization, power-law spectrum, and the presence of spectral index gradient in the downstream regions, we classify it as a radio relic. R1 morphology is arc-shaped and it is more extended toward low frequencies. Given its peripheral location, spectral properties, the radio power vs. LLS relation, and the presence of a density jump, we conclude that this source is also a radio relic.  

R3 is located symmetrical to the NW relic, suggesting that it could be a double radio relic system. The location of R3, its megaparsec size, the radio power vs. LLS relation, and the detected surface brightness jump across the sector SB3 suggest that it could be a relic. However, its integrated spectral index is inconsistent with this since in the case of the shock acceleration a power-law spectrum is expected while the R3 spectrum exhibits a high frequency steepening. Low-frequency observations (LOFAR 50 MHz and uGMRT band3) might allow an accurate measurement of the spectral index distribution. If R3 shows a radial spectral index steepening that would favor shock re-acceleration and the complex spectrum can be explained that the low-frequency spectrum marks the original slope of the fossil seeds electrons and the high frequency would represent the power-law of DSA (re-acceleration). The nature of R2 is not clear since it is only detected at 150\,MHz.


\section{Summary and conclusions}
\label{summary}

In this work, we presented deep \textit{XMM-Newton} ($\sim$150\,ks), uGMRT Band4 (550-750\,MHz) and VLA L-band (1-2\,GHz) observations of the complex galaxy cluster Abell 746. The main aims of our analysis were to identify signs of merger activity in the ICM and to detect and characterize diffuse radio sources. Below is a summary of our main results:
\begin{enumerate}
\item{}  The \textit{XMM-Newton} images reveal a complex ICM distribution in Abell 746, which includes a V-shaped feature, an elongated central region, and concave bay and arc-like features. We find evidence of asymmetric temperatures across the cluster ranging from ($\leq 4 \rm \,keV$)  in the north and $\sim9\,\rm keV$ in the south. 

\item{} We detected three surface brightness edges and one candidate edge. Three of these are merger-driven shock fronts located in the southeast, southwest, and north of the cluster.

\item{} Abell 746 possibly hosts double radio relics, in addition to two fainter relics to the east (R1 and R2) and a radio halo with low surface brightness.

\item{}  The main northwest relic shows filamentary morphology. We measure a power law spectrum, a clear spectral index gradient, and a high degree of polarization, as expected for acceleration from an outward traveling shock, with radiative losses in the region downstream regions. The relic shows a single RM component with a mean RM ($-10\rm\,rad\,m^{-2}$) very close to the Galactic foreground, indicating very little Faraday-rotating intervening material along the line of sight.

\item{}  The integrated spectral index of the fainter northern relic R1 is  $\alpha_{\rm 144\,MHz}^{\rm 1.5\,GHz}=-1.36\pm0.07$, implying a shock of Mach number $\mathcal{M}=2.4\pm0.2$. This is consistent with X-ray obtained shock Mach number of $\mathcal{M}=2.6\pm0.8$.

\item{}  R3 shows a high-frequency steepening in the overall spectrum. The ICM temperature across a part of R3 is high, consistent with shock heating.  We find evidence of a density jump in this region.

\item{} Abell 746 is host to a 1.7\,Mpc large radio halo with low surface brightness.  The integrated spectrum of the halo follows a power law and has a slope of $\alpha_{\rm 144\,MHz}^{\rm 1.5\,GHz}=-1.48\pm0.10$. The radio power of the halo at 1.4 GHz and 144 MHz is consistent with the $P-M_{500}$ and $P-L_{X}$ relations of known radio halos. Unlike other known halos, we do not find a morphological connection between the radio halo emission and the thermal X-ray emission. 
\end{enumerate}

While progress has been made in understanding particle acceleration and the origin of non-thermal emission in galaxy clusters through both radio and X-ray observations, questions remain. Multiwavelength observations combined with simulations will shed light on the underlying particle acceleration mechanisms and the complex interplay between thermal and nonthermal plasma. In particular, sensitive low frequencies radio observations, such as those that can be obtained with the LOFAR Low Band Antennas (LBA) will be ideal to probe the existence and spatial distribution of older electron populations. Finally, \textit{Chandra} high resolution observations may allow to characterize the localized regions and investigate the connection between the radio features and the detected discontinuities.

\section*{Acknowledgments}
We thank the anonymous referee for a constructive report. KR, WF, and CJ acknowledge support from the Smithsonian Institution and NASA  80NSSC22K1621. KR and FV acknowledges ERC starting grant ``MAGCOW" No. 714196. LL acknowledges financial contribution from the INAF grant 1.05.12.04.01. RJvW acknowledges support from the ERC Starting Grant ClusterWeb 804208. AB acknowledges support from the ERC through grant ERC-Stg DRANOEL No. 714245. M. J. Jee acknowledges support for the current research from the National Research Foundation (NRF) of Korea under the programs 2022R1A2C1003130 and RS-2023-00219959. PDF acknowledges the support of the Future Faculty Leaders Fellowship. AS acknowledges the support of a Clay Fellowship. This work is based on observations obtained with XMM-Newton, an ESA science mission with instruments and contributions directly funded by ESA Member States and the US (NASA). The National Radio Astronomy Observatory is a facility of the National Science Foundation operated under cooperative agreement by Associated Universities. The GMRT is run by the National Centre for Radio Astrophysics (NCRA) of the Tata Institute of Fundamental Research (TIFR).  LOFAR \citep{Haarlem2013} is the Low Frequency Array designed and constructed by ASTRON. It has observing, data processing, and data storage facilities in several countries, which are owned by various parties (each with their own funding sources), and that are collectively operated by the ILT foundation under a joint scientific policy. The ILT resources have benefited from the following recent major funding sources: CNRS-INSU, Observatoire de Paris and Universit\'{e} d'Orl\'{e}ans, France; BMBF, MIWF-NRW, MPG, Germany; Science Foundation Ireland (SFI), Department of Business, Enterprise and Innovation (DBEI), Ireland; NWO, The Netherlands; The Science and Technology Facilities Council, UK; Ministry of Science and Higher Education, Poland; The Istituto Nazionale di Astrofisica (INAF), Italy. This research made use of the LOFAR-UK computing facility located at the University of Hertfordshire and supported by STFC [ST/P000096/1], and of the LOFAR-IT computing infrastructure supported and operated by INAF, and by the Physics Dept. of Turin University (under the agreement with Consorzio Interuniversitario per la Fisica Spaziale) at the C3S Supercomputing Centre, Italy. 

\bibliography{ref.bib}

\end{document}